\pdfoutput=1
\documentclass[11pt,table]{article}

\usepackage[final]{acl}

\usepackage{times}
\usepackage{latexsym}
\usepackage{microtype}
\usepackage{graphicx}
\usepackage{booktabs} % for professional tables
\usepackage{multirow,tabularx,multicol}
\definecolor{verylightgray}{gray}{0.9}
\definecolor{verylightred}{rgb}{1,0.9,0.9}
\usepackage{hyperref}
\usepackage{subcaption}

\usepackage{amsmath}
\usepackage{amssymb}
\usepackage{mathtools}
\usepackage{amsthm}
\usepackage{amsfonts}

\usepackage{paralist}

\usepackage[T1]{fontenc}
\usepackage[utf8]{inputenc}

\usepackage{microtype}

\usepackage{inconsolata}

\usepackage{graphicx}

\title{OZSpeech: One-step Zero-shot Speech Synthesis with Learned-Prior-Conditioned Flow Matching}

\author{
 \textbf{Hieu-Nghia Huynh-Nguyen\textsuperscript{1}},
 \textbf{Ngoc Son Nguyen\textsuperscript{1}},
 \textbf{Huynh Nguyen Dang\textsuperscript{1}},\\
 \textbf{Thieu Vo\textsuperscript{1}},
 \textbf{Truong-Son Hy\textsuperscript{2}},
 \textbf{Van Nguyen\textsuperscript{1}},
\\
 \textsuperscript{1}FPT Software AI Center, Vietnam \\
 \texttt{\{nghiahnh, sonnn45, huynhnd11, thieuvn2, vannth19\}@fpt.com} \\
 \textsuperscript{2}University of Alabama at Birmingham, USA \\
 \texttt{thy@uab.edu} \\
}

\begin{document}
\maketitle

\begin{abstract}
Text-to-speech (TTS) systems have seen significant advancements in recent years, driven by improvements in deep learning and network architectures. Viewing the output speech as a data distribution, previous approaches often employ traditional speech representations, such as waveforms or spectrograms, within the Flow Matching framework. However, these methods have limitations, including overlooking various speech attributes and incurring high computational costs due to additional constraints introduced during training. To address these challenges, we introduce \textit{OZSpeech}, the first TTS method to explore optimal transport conditional flow matching with one-step sampling and a learned prior as the condition, effectively disregarding preceding states and reducing the number of sampling steps. Our approach operates on disentangled, factorized components of speech in token format, enabling accurate modeling of each speech attribute, which enhances the TTS system's ability to precisely clone the prompt speech. Experimental results show that our method achieves promising performance over existing methods in content accuracy, naturalness, prosody generation, and speaker style preservation. Code and audio samples are available at our demo page \footnote{\url{https://ozspeech.github.io/OZSpeech_Web/}}.

% Text-to-speech (TTS) systems have seen significant advancements in recent years, driven by improvements in deep learning and neural network architectures. This paper introduces \emph{OZSpeech}, the first TTS method that explores optimal transport conditional flow matching with one-step sampling with a learned prior as the condition. Instead of using traditional speech representations such as waveforms or spectrograms, this method operates on disentangled factorized components of speech in token format. Experiment results show that our method achieves superior performance over existing methods in content accuracy, naturalness, prosody generation, and speaker style preservation. Audio samples are available at our demo page \footnote{\url{https://ozspeech.github.io/OZSpeech_Web/}}.

%We discuss the key challenges in the field, including naturalness, prosody generation, and speaker variability, and examine how recent innovations such as Tacotron, WaveNet, and FastSpeech address these issues. Additionally, we present an overview of the datasets, evaluation metrics, and performance benchmarks commonly used in TTS research, highlighting their limitations and areas for improvement. Finally, the paper investigates the future of TTS, including its integration with multimodal systems, applications in assistive technologies, and potential ethical considerations surrounding the technology’s use. 

\end{abstract}

\section{Introduction}
Text-to-speech (TTS) has numerous real-world applications, such as voice-based virtual assistants, assistive screen readers for the visually impaired, and reading aids for people with dyslexia, to name a few. Most TTS systems focus on synthesizing speech that matches a speaker in a set of speakers seen during training. Recent studies tackle a more challenging problem of converting text into speech that follows the acoustic characteristics of a prompt spoken by a speaker not seen during training. This problem is called zero-shot TTS.

% Existing challenges in TTS that researchers are still facing include lack of similarity with target speaker, lack of punctuation, unnatural/robotic synthesized speech, and long inference time. Long inference time is particularly problematic for time-sensitive applications such as virtual voice assistants and virtual or augmented reality, as it significantly degrades the user experience.

In recent years, remarkable progress has been achieved in the research of Zero-shot TTS models. 
These advancements have demonstrated the impressive capabilities of such models, with their synthesized outputs often approaching a quality level that is virtually indistinguishable from human speech. 
The body of research on Zero-Shot TTS can be broadly divided into two primary categories, each aligned with a dominant methodological paradigm in the field: autoregressive models and diffusion-based models.

Prominent examples of the autoregressive approach are VALL-E \cite{valle} and its variants \cite{valle2, valle-x, valle-r, melle, ella-v, voicecraft, mobilespeech}, which have significantly advanced Zero-Shot TTS by integrating language modeling techniques and employing disentangled speech units as input and output tokens. 
This innovative framework has paved the way for the potential convergence of Zero-Shot TTS with large language models (LLMs), enabling the creation of efficient, multimodal systems which are capable of generating text, speech, and other modalities in a flexible and scalable manner. 
However, as with other LLM-based systems, autoregressive models are susceptible to the issue of the non-deterministic sampling process, potentially leading to infinite repetition, which remains a critical challenge in applications requiring high levels of precision and reliability.

In contrast, diffusion-based models, as demonstrated by state-of-the-art (SOTA) TTS systems such as E2 TTS \cite{e2tts} and other related approaches \cite{voicebox, audiobox, ns2, ns3}, have emerged as powerful generative frameworks capable of producing high-quality, natural-sounding audio.
This approach has proven particularly effective in specialized tasks such as in-filling and speech editing. 
Nevertheless, diffusion-based models face limitations in real-time applications due to the computational inefficiency of their multi-step sampling processes. 
%These constraints highlight the trade-offs inherent in the diffusion-based paradigm when applied to scenarios necessitating low-latency performance.
These constraints underscore the trade-offs inherent in the diffusion-based paradigm, particularly in scenarios that demand low-latency performance.

Distillation methods for diffusion-based models have been explored to address the multi-step sampling challenge, with Consistency Models \cite{consistencymodels} introducing one-to-one mapping functions that transform intermediate states along the Ordinary Differential Equation (ODE) trajectory directly to their origin. 
This approach reduces sampling steps to one while maintaining output quality but requires access to a full range of $t \in [0, 1]$ to approximate trajectories, demanding extensive training steps. 
As an alternative, Shortcut Models \cite{shortcutmodels} condition the network on noise level and step size, enabling faster generation with fewer training steps by using only a subset of $t$ values. 
However, this method is computationally intensive due to additional constraints introduced during training, making it more resource-demanding than Consistency Models.

To capitalize on the strengths and mitigate the limitations of the aforementioned approaches, we propose OZSpeech (\textbf{O}ne-step \textbf{Z}ero-shot \textbf{Speech} Synthesis with Learned-Prior-Conditioned Flow Matching), a novel Zero-Shot TTS system.
Our model leverages optimal transport conditional flow matching \cite{ot-cfm} (OT-CFM), a class of diffusion-based models. 
We reformulate the original OT-CFM to enable single-step sampling, where the vector field estimator regresses the trajectories of all pairs of initial points from the learned prior distribution, rather than conventional Gaussian noise, to their respective target distributions.
By minimizing the distance between the initial points and their origins while implicitly learning the optimal $t$ for each prior, this approach eliminates the need to access a comprehensive range of $t$ values or compute additional constraints, thereby ensuring high-fidelity synthesized speech.

The key contributions of this paper are as follows:
\begin{compactitem}
%\begin{enumerate}   
    \item We propose a reformulated OT-CFM framework that effectively initializes the starting points of the flow matching process using samples from a learned prior distribution. This prior is optimized to closely approximate the target distribution, enabling one-step sampling with minimal errors. Our framework requires only a single training run without the need for an extensive distillation stage.
    
    \item We propose a simple yet effective network architecture to learn prior-distributed codes.
        
    \item Compared to previous methods, our model yields multi-fold improvement in WER and latency, achieving significant reduction in model size while striking a balance with acoustical quality. In addition, while previous models suffer from increasing noise level in the audio prompts, OZSpeech's WER remains stable, highlighting the excellent noise-tolerant intelligibility of our method. Our model requires significantly less computation, with inference speed being $2.7 - 6.5$ times faster than the other methods. Our model is only 29\%-71\% the size of the other models.
    % Our model outperforms SOTA methods on a range of objective evaluation metrics yet have significantly smaller size and shorter inference time. We achieve model size reduction of 76\% and faster inference time by 70\% compared to SOTA methods, respectively.

%\end{enumerate}
\end{compactitem}

% data preprocessing different rom existing works: input truc tiep vao models are not waveforms or mel but codec quantizers. Such dataset wil be published

%%%%%%%%%%%%%%%%%%%%%%%%%%%%%%%%%%%%%%%%%%%%%%%%%%%%%%%%%%%%%%%%%%%%%%%%%%%%%%%%%%%%%%%%%%%
%BACKGROUND & PRELIMINARIES
%%%%%%%%%%%%%%%%%%%%%%%%%%%%%%%%%%%%%%%%%%%%%%%%%%%%%%%%%%%%%%%%%%%%%%%%%%%%%%%%%%%%%%%%%%%
% \section{Related Work \& Preliminaries}

% Related Work:

% - zero-shot tts \& neural codec

% Preliminaries:

% - FACodec

% - flow matching
\section{Related Work}
% \subsection{Zero-shot TTS \& Neural Codec}
Zero-Shot TTS enables the generation of speech in an unseen speaker's voice using only a few seconds of audio as a prompt; this process is often termed voice mimicking. Advances in large-scale generative models have driven significant progress in this field. One prominent development is the adoption of diffusion models \cite{NEURIPS2020_4c5bcfec, song2021scorebased}, which have demonstrated remarkable performance \cite{kang23_interspeech, tran23d_interspeech, ns2, ns3}. Another approach, flow matching \cite{ot-cfm, rectified-flow}, has further advanced the state-of-the-art by delivering strong results with reduced inference times \cite{kim2023pflow, matcha-tts, e2tts, f5tts}. Additionally, a key innovation in Zero-Shot TTS is the use of discrete tokens, often derived from neural codecs \cite{wang2023neural, kharitonov-etal-2023-speak, chen2024vall, ju2024naturalspeech, unicasts}.

Neural codecs are designed to learn discrete speech and audio tokens, often referred to as acoustic tokens, while preserving reconstruction quality and maintaining a low bitrate. SoundStream \cite{soundstream} is a well-known example that employs a vector-quantized variational autoencoder (VQ-VAE), which was first introduced by \cite{VQ-VAE} in the field of computer vision, and later adapted to TTS, to disentangle continuous data into discrete tokens.
It comprises multiple residual vector quantizers to compress speech into multiple tokens, which serve as intermediate representations for speech generation. A significant breakthrough in this area, inspired by the success of LLMs in natural language processing, is VALL-E \cite{valle}, a pioneering work in this domain. VALL-E represents speech as discrete codec codes using an off-the-shelf neural codec and redefines TTS as a conditional codec language modeling task. This approach has sparked further research and development in the field \cite{kharitonov-etal-2023-speak, valle-x, valle2, valle-r, du2024cosyvoice}.

\section{Method}

\begin{figure*}[htbp]
    \vspace{-0.5cm}
    \centering
    \includegraphics[width=1\textwidth]{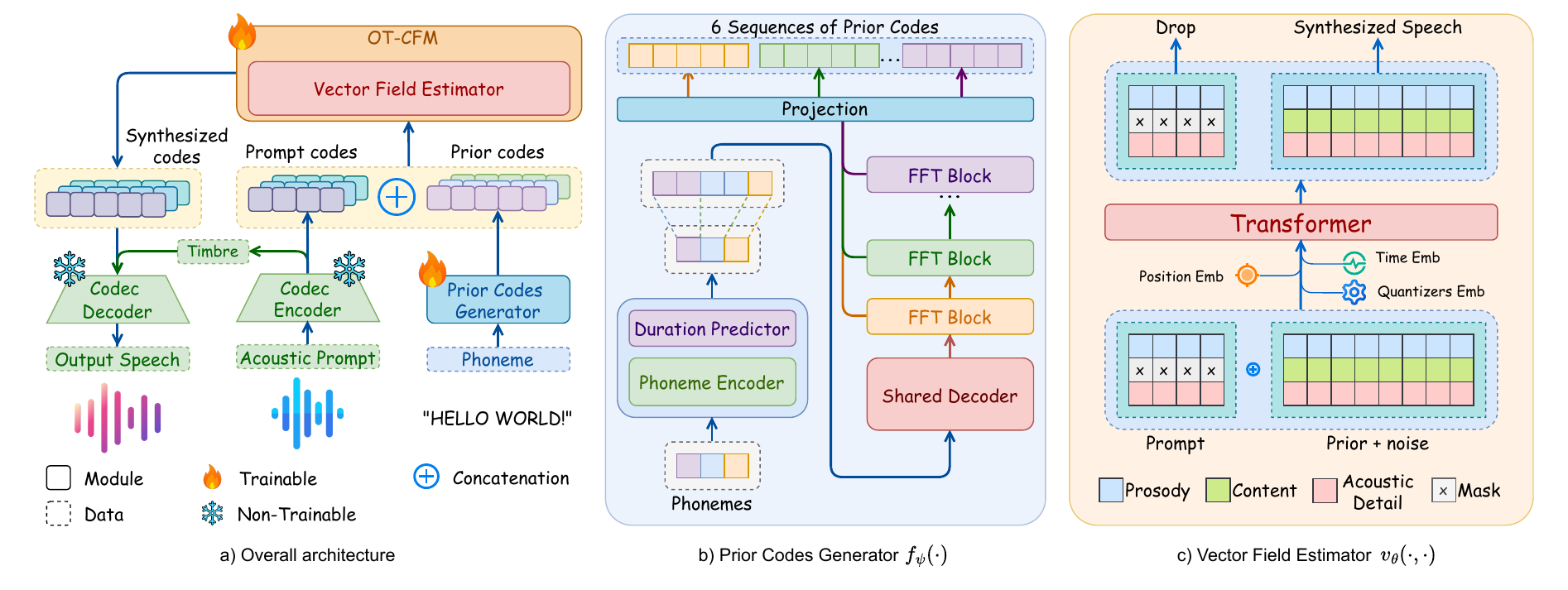}
    \caption{Overview of OZSpeech: (a) The overall architecture: The text prompt is converted to phonemes and then into prior codes via the Prior Codes Generator. Simultaneously, the audio prompt is encoded into codes using the FACodec Encoder. These codes are concatenated along the sequence dimension and fed into the OT-CFM Vector Field Estimator, which generates codes preserving the text content and acoustic attributes. Finally, the FACodec Decoder converts them into output speech. (b) The Prior Codes Generator $f_\psi(\cdot)$ produces sequences of phoneme-aligned codes. 
    (c) The Vector Field Estimator refines these codes with the prosody and acoustic details from the acoustic prompt. Before being fed through $v_\theta(\cdot, \cdot)$, six sequences of codes are first enhanced via Quantizer Embedding, which serves as an identifier for each sequence within the hidden space. These embeddings are then folded along the hidden dimension and processed by the network to estimate the velocity of the prior codes. 
    % The input sequence is subsequently enriched with Position Embedding and Time Embedding before being processed by the network to estimate the velocity of the prior codes.
    }
    \label{fig:overall-architecture}
    \vspace{-0.5cm}
\end{figure*}

\subsection{Problem Statement}
\label{sec:problem-statement}

In this paper, we consider the problem of generating speech from given text and acoustic prompt such that conditions for the outputs are met. 
Viewing the synthesized speech as a data distribution, denoted as $\text{x}_1 \sim p_1(x)$, previous methods often construct the output data distribution from a noise distribution $\text{x}_0 \sim p_0(x)$. 
However, we propose constructing the output data distribution from a feasible intermediate state candidate $\text{x}_{\text{pr}} \sim p_{\text{prior}}(x)$ instead of $\text{x}_0$, thereby disregarding preceding states and reducing the number of sampling steps. 
To achieve this, we undertake the following steps:
\begin{compactitem}
    \item \textbf{Prior Code Generation:} We design an effective method for generating prior codes to produce $\textbf{x}_{\text{pr}}$ (see Section \ref{sec:prior-codes-gen}).
    \item \textbf{Vector Field Estimation:} We develop a vector field estimator to approximate $v_{\theta}$ , facilitating the transition from $\text{x}_{\text{pr}}$ to $\text{x}_{1}$ (see Section \ref{sec:osfm}).
    \item \textbf{Waveform Decomposition via FACodec:} We employ FACodec \cite{ns3}, a neural codec disentangler framework, to decompose the waveform into distinct components, including speaker identity and sequences of codes encoding prosody, content, and acoustic 
  details. This decomposition enables precise control over the aspects of speech to be preserved or modified.
\end{compactitem}

\subsection{Prior Codes Generation Modeling}
\label{sec:prior-codes-gen}
% In previous studies utilizing autoregressive (AR) architectures as generative models \textemdash such as VALL-E \cite{valle}, MELLE \cite{melle}, ELLA-V \cite{ella-v}, and VoiceCraft \cite{voicecraft} \textemdash a prevalent challenge has been the inherent non-deterministic sampling algorithm. 
% The algorithm often leads to critically repetitive outputs, thereby compromising the overall effectiveness of the models.
% To address this issue, several works have adopted fully non-autoregressive (NAR) architectures as their foundational framework. 
% Notable examples include Voicebox \cite{voicebox}, \textit{NaturalSpeech2} \cite{ns2}, \textit{NaturalSpeech3} \cite{ns3}, and E2 TTS \cite{e2tts}.
% Leveraging the advantages of NAR models, we propose a fully NAR neural network architecture specifically designed to approximate discrete quantizers of speech. 
% This architecture aims to establish a prior distribution based on the provided text input, thereby enhancing the generative capabilities of speech synthesis.
% In our framework, we employ the Feed-Forward Transformer (FFT), as introduced by \cite{fs}, to implement our architecture.

Our key contribution to prior code generation is that the process follows a hierarchical structure: each code sequence generation depends on the preceding code sequences, while the condition for the first code sequence is initialized based on phoneme embeddings.
To achieve this, we implement a cascaded neural network where specific decoder layers generate the respective code sequences in the hierarchy (shown in Fig.~\ref{fig:overall-architecture}b).
Formally, the Prior Codes Generator $\mathbf{f}_{\psi}(\cdot)$ is modeled as:
\begin{equation}
    \label{eq:pcg}
    p(\mathbf{q}_{1:6} \mid \mathbf{p}; \psi) = p(\mathbf{q}_{1} \mid \mathbf{p}; f_{\psi}^1) \prod_{j=2}^{6} p(\mathbf{q}_{j} \mid \mathbf{q}_{j-1}; f_{\psi}^j),
\end{equation}
% Here, $\mathbf{q}_j$ denotes the $j$-th quantizer produced at the same level of the FFT decoder layer $f_{\psi}^j$, while $\mathbf{p}$ represents the phoneme embeddings serving as the initial condition.
% The Prior Loss, $\mathcal{L}_{\text{prior}}$, is employed to minimize the negative log-likelihood of the joint probability, as defined in Equation \ref{eq:pcg}.
% This approach ensures effective learning of the content quantizers by conditioning on input phonemes while the other quantizers converge toward the mean representations of attributes such as pitch, energy, and acoustic details.
% In this framework, the Prior Codes Generator is designed to produce quantizers that are semantically meaningful to some extent. 
% In other words, it minimizes the distance between $\mathbf{x}_{\text{pr}}$ and $\mathbf{x}_1$, enabling the Vector Field Estimator $\mathbf{v}_{\theta}(\cdot, \cdot)$ to approximate vectors starting from a pre-determined distribution (in this case, the mean distribution) rather than generating them from scratch.

\noindent where \(\mathbf{q}_j\) is the \(j\)-th code sequence from the Feed-Forward Transformer (FFT) decoder layer \(f_{\psi}^j\) and \(\mathbf{p}\) represents phoneme embeddings as the initial condition. 
The Prior Loss \(\mathcal{L}_{\text{prior}}\) minimizes the negative logarithm of the joint probability in Eq.~ \eqref{eq:pcg}, ensuring content code sequences are learned effectively by conditioning on phonemes, while the others, including prosody and acoustic details, converge towards the mean representations.
The Prior Codes Generator produces semantically meaningful codes, reducing the distance between \(\mathbf{x}_{\text{pr}}\) and \(\mathbf{x}_1\), allowing the Vector Field Estimator \(\mathbf{v}_{\theta}(\cdot, \cdot)\) to approximate vectors from a mean distribution rather than generating them from pure noise.

To align the input phonemes with their corresponding output code sequences, we employ a neural network functioning as a Duration Predictor, as introduced in \cite{fs}. 
Briefly, the Duration Predictor estimates the duration (i.e., the number of acoustic tokens) for each input phoneme. 
The phoneme embeddings are duplicated accordingly before being passed through the decoder of the Prior Codes Generator.
We define the loss function used to train the Duration Predictor as $\mathcal{L}_{dur}$, which aims to minimize the mean squared error between the predicted and ground truth durations on a logarithmic scale.
% The objective of this module is formulated as follows:
% \begin{equation}
%     \label{eq:dur}
%     \mathcal{L}_{dur}(\delta) = \mathbb{E}_{d \sim \mathcal{D}}
%     \left\lVert
%         \log d - \log \tilde{d}
%     \right\rVert^2,
% \end{equation}
% where $d$ and $\tilde{d}$ denote the ground truth duration and the estimated duration produced from a $\delta$-parameterized duration predictor, respectively.

\subsection{One-Step Optimal Transport Flow Matching for Zero-Shot TTS}
\label{sec:osfm}

\textbf{One-Step Optimal Transport Flow Matching Formulation}.
We rectify the OT-CFM paradigm, simultaneously introduced by \cite{ot-cfm} and \cite{rectified-flow} and first adapted to text generation by \cite{flowseq}.
Our approach involves constructing a vector field that regresses the velocity of non-Gaussian distribution and data distribution pairs, where the initial distribution closely approximates the target distribution.
To this end, we reformulate the original flow matching loss equation (details provided in \ref{appendix:fm}) to explicitly account for the discrepancies between the non-Gaussian initial distribution and the target distribution. Let $\mathbf{x}_t$ denote a linear probability path, starting from a purely noisy initial point $\mathbf{x}_0 \sim \mathcal{N}(0, I)$ and progressing towards a data point $\mathbf{x}_1 \sim \mathcal{D}$. It is defined as $\mathbf{x}_t = t\mathbf{x}_1 + (1-t)\mathbf{x}_0$,
% \begin{equation}
%     \label{eq:ot}
%     \mathbf{x}_t = t\mathbf{x}_1 + (1-t)\mathbf{x}_0,
% \end{equation}
where $t \sim \mathcal{U}(0, 1)$ denotes the interpolation parameter. 
Based on this, the initial point $\mathbf{x}_0$ can be derived as $\mathbf{x}_0 = \frac{t\mathbf{x}_1 - \mathbf{x}_t}{1 - t}$. 
Thus, the OT-CFM objective, as presented in Eq.~\eqref{eq:orgin-cfm}, can be reformulated as follows:
\begin{equation}
    \label{eq:ot-cfm-adjusted}
    \mathcal{L}_{CFM}(\theta) = \mathbb{E}_{t,\mathbf{x}_0,\mathbf{x}_1}
    \left\lVert
        \mathbf{v}_\theta(\mathbf{x}_t, t) - \frac{\mathbf{x}_1 - \mathbf{x}_t}{1 - t}
    \right\rVert^2.
\end{equation}
We assume that $\mathbf{x}_t$ can be estimated via $\mathbf{f}_{\psi}(\cdot)$. Under this assumption, $\mathbf{x}_t$ is treated as a learnable state, which we denote as $\mathbf{x}_{\text{pr}}$, while $t$ is regarded as an unknown interpolation parameter, represented as a prior-dependent time variable $\tau$. 
Consequently, Eq.~(\ref{eq:ot-cfm-adjusted}) can be reformulated as follows:
\begin{equation}
    \label{eq:ot-cfm-adjusted-1}
    \begin{split}
        \mathcal{L}_{\text{CFM}}(\theta) = \mathbb{E}_{\mathbf{x}_{\text{pr}}, \mathbf{x}_1}
        \left\lVert
            \mathbf{v}_\theta(\mathbf{x}_{\text{pr}}, \tau) - \frac{\mathbf{x}_1 - \mathbf{x}_{\text{pr}}}{1 - \tau}
        \right\rVert^2.
    \end{split}
\end{equation}
This paradigm is similar to the original OT-CFM in its objective of regressing the velocity between the initial prior $\mathbf{x}_0$ and data $\mathbf{x}_1$ pairs. 
However, unlike the original approach, it does not access the distribution of $\mathbf{x}_0$ during training.
Therefore, it also does not enforce $\mathbf{x}_0$ to follow a normal distribution.
Furthermore, as the prior distribution $\mathbf{x}_{\text{pr}}$ approaches the target distribution $\mathbf{x}_1$, both the number of sampling steps and the magnitude of each step are substantially reduced. 
This convergence enables the sampling process to be efficiently performed in as few as a single step.\\[8pt]
\textbf{Vector Field Estimator Modeling}. 
To model $\mathbf{v}_{\theta}(\cdot,\cdot)$, we define the latent representations of the prior, target, and estimated target distributions as $\mathbf{x}_{\text{pr}}$, $\mathbf{x}_1$, and $\mathbf{\tilde{x}}_1$, respectively, where $\mathbf{x}_{\text{pr}}, \mathbf{x}_1, \mathbf{\tilde{x}}_1 \in \mathbb{R}^{6 \times N \times D}$. 
These terms represent six quantizer categories with sequence length $N$ and feature dimensionality $D$. 
Inspired by diffusion-based text generation models like Diffuseq \cite{diffuseq}, Difformer \cite{difformer}, and FlowSeq \cite{flowseq}, we condition $\mathbf{x}_{\text{pr}}$ on an acoustic prompt to guide $\mathbf{v}_{\theta}(\cdot,\cdot)$ in transferring speech attributes other than content.
% The Prior Codes Generator $\mathbf{f}{\psi}(\cdot)$ synthesizes speech with consistent prosody and acoustic detail, emphasizing the role of acoustic prompts in preserving speech attributes.
Specifically, only prosody and acoustic detail codes serve as prompts, while the content codes are masked to prevent undesired content transfer.
% Let $\mathbf{y} = \{y_{i} \mid i=1\dots6\}$, $y_{i} \in \mathbb{R}^{M \times D}$, denotes the acoustic prompt, where $M$ corresponds to the sequence length of the prompt.
% The content-masked acoustic prompt is defined as follows:
% \begin{equation}
%     \label{eq:y-mask}
%     \mathbf{y}_{\text{mask}} = \left\{
%     \begin{array}{ll}
%         y_i \odot \mathbf{0}^{M \times D}, & \text{if } i \in \{2, 3\}, \\
%         y_i \odot \mathbf{1}^{M \times D}, & \text{otherwise}.
%     \end{array}
%     \right.
% \end{equation}
The input to $\mathbf{v}_{\theta}(\cdot,\cdot)$ is then formulated as the concatenated representation:
\begin{equation*}
    \label{eq:z-prior}
    \mathbf{z_{\text{pr}}} = \textbf{Concat}(\mathbf{y}_{\text{mask}}, \mathbf{x}_{\text{pr}} + \epsilon), \quad  \epsilon \sim \mathcal{N}(0,I),
\end{equation*}
where $\mathbf{z}_\text{pr} \in \mathbb{R}^{6 \times L \times D}$, with $L = M + N$ representing the total sequence length of the concatenated input.
$\mathbf{y}_\text{mask}$ denotes the content-masking representation of the acoustic prompt.
Random Gaussian noise $\epsilon$ is added to the latent representation of $\mathbf{x}_{\text{pr}}$ to ensure the robustness and diversity of the model.
The diagram of the Vector Field Estimator is shown in Fig.~\ref{fig:overall-architecture}c.\\[8pt]
\textbf{Folding Mechanism and Quantizer Encoding}.
% Our data representation consists of six sequences of quantizer embeddings, making it unsuitable for direct sequence modeling with Transformer architectures.
% To address this challenge, several prior works employing Neural Codec to discretize audio, such as VALL-E \cite{valle}, VALL-E 2 \cite{valle2}, and particularly \textit{NaturalSpeech3}  \cite{ns3}, have opted to model each quantizer sequence independently. 
% While effective to some extent, this approach suffers from significant limitations, including high computational cost and extended generation times, as each quantizer is processed sequentially.
% Additionally, in the case of \textit{NaturalSpeech3}, the inference time becomes prohibitively long due to the diffusion-based sampling procedure, further impacting its practicality for real-time or large-scale applications.
% To address this issue, we propose a methodology that integrates the modeling of all six quantizers simultaneously by organizing them along the hidden dimension in a folded configuration.
% Let $\mathcal{F}(\cdot)$ denote the folding function, which can be formally defined as a composition of two functions: $\mathcal{G}: \mathbb{R}^{6 \times L \times D} \to \mathbb{R}^{L \times 6D}$ and $\mathcal{H}: \mathbb{R}^{L \times 6D} \to \mathbb{R}^{6D \times D'}$. Thus, $\mathcal{F}$ can be expressed as:
Our data representation consists of six sequences of quantizer embeddings, making direct sequence modeling with Transformers challenging. 
Previous works using Neural Codec for audio discretization, such as VALL-E \cite{valle}, VALL-E 2 \cite{valle2}, and \textit{NaturalSpeech3} \cite{ns3}, model each quantizer sequence independently. 
While effective, this approach incurs high computational costs and long generation times due to sequential processing.
% \textit{NaturalSpeech3} further suffers from slow inference due to its diffusion-based sampling, limiting its practicality.
To mitigate these inefficiencies, we propose modeling all six quantizers simultaneously by folding them along the hidden dimension. 
Let $\mathcal{F}(\cdot)$ be the folding function, defined as a composition of two transformations: $\mathcal{G}: \mathbb{R}^{6 \times L \times D} \to \mathbb{R}^{L \times 6D}$ and $\mathcal{H}: \mathbb{R}^{L \times 6D} \to \mathbb{R}^{L \times D'}$. Thus, $\mathcal{F}$ is expressed as:
$\mathcal{F} \triangleq \mathcal{H} \circ \mathcal{G}: \mathbb{R}^{6 \times L \times D} \to \mathbb{R}^{L \times D'}.
$
Within this framework, the function $\mathcal{F}(\cdot)$ permutes and reshapes the input tensor to sequentially align the component quantizers in the hidden space, following the prescribed order of prosody, content, and acoustic details. 
Subsequently, it adjusts the tensor's dimensionality accordingly.
In addition, we propose a quantizer encoding mechanism designed to identify specific quantizers within the hidden space. 
This mechanism operates in conjunction with the $\mathcal{F}(\cdot)$ function. 
The quantizer encoding is formally defined as follows:
\begin{equation*}
    \label{eq:vq-encode}
    \mathcal{Q}(x) = x + \textbf{Dup}(\omega, L), \quad x \in \mathbb{R}^{6 \times L \times D}.
\end{equation*}
Here, the quantizer encoding function $\mathcal{Q}(\cdot)$ integrates the input tensor $\mathbb{R}^{6 \times L \times D}$ with the embedding $\textbf{Dup}(\omega, L)$.
In this context, $\omega \in \mathbb{R}^{6 \times 1 \times D}$ plays a role as the identifier for the latent representation of the quantizers, while the function $\textbf{Dup}(\cdot, \cdot)$ duplicates the identifier along the sequence length $L$.
With $\mathcal{Q}(\cdot)$, we aim to prevent the model from confusing the quantizers with each other, even when they are simultaneously modeled within a single sequence.
Consequently, the latent representation $\mathbf{z}_{\text{pr}}$ is transformed into $\breve{\mathbf{z}}_{\text{pr}}$, which subsequently serves as the direct input to the function $\mathbf{v}_{\theta}(\cdot, \cdot)$.
The transformation is defined as $\breve{\mathbf{z}}_{\text{pr}} = (\mathcal{F} \circ \mathcal{Q})(\mathbf{z}_{\text{pr}})$.\\[8pt]
% \begin{equation}
%     \label{eq:z-fold}
%     \breve{\mathbf{z}}_{\text{pr}} = (\mathcal{F} \circ \mathcal{Q})(\mathbf{z}_{\text{pr}}).
% \end{equation}
\textbf{Anchor Loss}.
% To achieve the best performance of our generative model on discrete token data as well as stabilize the training process, we incorporate the Anchor Loss $\mathcal{L}_{anchor}$ as a regularization term for the embeddings.
% Originally introduced in the diffusion models interacting with discrete data (e.g., text) by \cite{difformer} and later adapted to the flow matching models by \cite{flowseq}, the $\mathcal{L}_{anchor}$, which measures the difference between arbitrary intermediate state $\mathbf{x}_t$ the ground truth $\mathbf{x}_1$, has demonstrated effectiveness in preventing the embeddings from collapsing and mitigating the distances between every intermediate state $\mathbf{x}_{t}$ and their target $\mathbf{x}_{1}$, enabling few-steps or even one-step sampling manner.
% In this study, the anchor loss, $\mathcal{L}_{\text{anchor}}$, is utilized to minimize the negative log-likelihood of the joint probability, as defined in Equation \ref{eq:anchor-prob}.
To optimize generative model performance on discrete token data and stabilize training, we use Anchor Loss \(\mathcal{L}_{anchor}\) as a regularization term for embeddings.
Initially introduced in diffusion models for discrete data \cite{difformer} and later adapted for flow matching models \cite{flowseq}, \(\mathcal{L}_{anchor}\) measures the difference between an intermediate state \(\mathbf{x}_t\) and the ground truth \(\mathbf{x}_1\). 
It prevents embedding collapse, reduces distances between states, and enables efficient sampling. 
In this study, \(\mathcal{L}_{\text{anchor}}\) minimizes the negative log-likelihood of the joint probability (Eq.~\eqref{eq:anchor-prob}).

Let $\mathbf{e}_{\phi}(\cdot)=[e_1,e_2,\dots,e_V] \in \mathbb{R}^{V \times D}$ denotes the embedding lookup function with the vocabulary size of $V$.
Given $\mathcal{A}^{6 \times L}=\{\alpha_{i,j} \mid i=1 \dots 6; j=1 \dots L\}$, where $\alpha_{i,j} \in \{1 \dots V \}$,  represents a quantizer element of an $\mathbf{x}_1$ with the sequence length of $L$, the embedding of $\mathcal{A}^{6 \times L}$ can be expressed as $\mathbf{e}_{\phi}(\mathcal{A}^{6 \times L})=\{\varepsilon_{i,j} \mid i=1 \dots 6; j=1 \dots L\}$, where $\varepsilon_{i,j} \in \mathbb{R}^D$.
Thus, the joint probability approximating the target distribution $\mathbf{z}_1$, conditioned on the estimated sequences $\tilde{\mathbf{z}}_{1}$, can be expressed as follows:
\begin{equation}
    \label{eq:anchor-prob}
    p(\mathbf{z}_1 \mid \tilde{\mathbf{z}}_1;\phi) = \prod_{i=1}^{6} \prod_{j=1}^{L} p(\varepsilon_{i,j} \mid \tilde{\varepsilon}_{i,j}; \mathbf{e}_{\phi}),
\end{equation}
% \begin{equation}
%     \label{eq:anchor}
%     \mathcal{L}_{anchor}(\phi) = \mathbb{E}_{q(\mathbf{x}_1)}
%     \left[
%         -\log p(\mathbf{x}_1 \mid \tilde{\mathbf{x}}_1;\mathbf{e}_{\phi})
%     \right]
% \end{equation}
where $\tilde{\mathbf{z}}_{1}$ is the approximation of $\mathbf{z}_{1}$, deduced using the triplet consisting of the estimated vector $\mathbf{v}_{\theta}(\breve{\mathbf{z}}_{\text{pr}}, \tau)$, the prior state $\mathbf{z}_{\text{pr}}$, and the corresponding interpolate parameter $\tau$.
The function $\mathcal{F}^{-1}$ denotes the reverse function of $\mathcal{F}$. This relationship is expressed as $\tilde{\mathbf{z}}_1 = \mathbf{z}_{\text{pr}} + (1 - \tau)\mathcal{F}^{-1}(\mathbf{v}_{\theta}(\breve{\mathbf{z}}_{\text{pr}}, \tau))$.\\[8pt]
%\begin{equation*}
%    \label{eq:x1-est}
%    \tilde{\mathbf{z}}_1 = \mathbf{z}_{\text{pr}} + (1 - \tau)\mathcal{F}^{-1}(\mathbf{v}_{\theta}(\breve{\mathbf{z}}_{\text{pr}}, \tau)).
%\end{equation*}
\textbf{Total loss}. We define the total loss function used in our joint training method as $\mathcal{L}_{total} = \mathcal{L}_{prior} + \mathcal{L}_{dur} + \mathcal{L}_{CFM} + \mathcal{L}_{anchor}$.
%\begin{equation*}
%\label{eq:loss-total}
%\mathcal{L}_{total} = \mathcal{L}_{prior} + \mathcal{L}_{dur} + \mathcal{L}_{CFM} + \mathcal{L}_{anchor}.
%\end{equation*}
%
The loss functions \(\mathcal{L}_{prior}\) and \(\mathcal{L}_{dur}\) set the training objective for the Prior Codes Generator, whereas \(\mathcal{L}_{CFM}\) and \(\mathcal{L}_{anchor}\) are designed to construct the vector field and distill the sampling steps, respectively.
\section{Experiments}

\begin{table*}
\caption{Performance evaluation on the \textit{LibriSpeech test-clean} across different audio prompt lengths. \textbf{Bold} indicates the best result, and \underline{underline} indicates the second-best result. ($\uparrow$) indicates that higher values are better, while ($\downarrow$) indicates that lower values are better. $[\spadesuit]$ means reproduced results. $[\bigstar]$ and $[\clubsuit]$ mean results inferred from official and ufficial checkpoints, respectively. Abbreviation: LT (LibriTTS), E (Emilia), GS (GigaSpeech).}
\label{tab:overall-results}
\vspace{-0.5cm}
\begin{center}
\resizebox{\textwidth}{!}{%
\begin{tabular}{lccccccccc}
\toprule
 & & & & \multicolumn{2}{c}{\textbf{SPK-SIM}} & \multicolumn{2}{c}{\textbf{F0}} & \multicolumn{2}{c}{\textbf{Energy}} \\
\cmidrule(lr){5-6} \cmidrule(lr){7-8} \cmidrule(lr){9-10 }
\textbf{Model} & \textbf{Data (hours)} & \textbf{UTMOS} ($\uparrow$) & \textbf{WER} ($\downarrow$) &\textbf{SIM-O} ($\uparrow$) & \textbf{SIM-R} ($\uparrow$) & \textbf{Accuracy} ($\uparrow$) & \textbf{RMSE} ($\downarrow$)& \textbf{Accuracy} ($\uparrow$)& \textbf{RMSE} ($\downarrow$) \\
\midrule
Ground Truth & & 4.09 & 0.02 & - & - & - & - & - & - \\
\midrule
% \rowcolor{verylightgray}
\multicolumn{10}{c}{\textit{1s Prompt}} \\
\midrule
% YourTTS $[\bigstar]$ \cite{yourtts} & LT + VCTK + T-P + M-A F & 640 & \textbf{3.61} & \underline{0.09} & \textbf{0.35} & - & 1.21 & \underline{1.36} \\
F5-TTS $[\bigstar]$ \cite{f5tts} & E (95,000) & \textbf{3.73} & 0.19 & \textbf{0.32} & - & 0.61 & 29.93 & \underline{0.50} & \underline{0.02} \\
VoiceCraft $[\bigstar]$ \cite{voicecraft} & GS (9,000) & 3.45 & 0.16 & \underline{0.31} & 0.24 & 0.61 & 31.57 & \textbf{0.52} & \textbf{0.01}\\
NaturalSpeech 2 $[\clubsuit]$ \cite{ns2} & LT (585) & 2.12 & \underline{0.12} & 0.20 & 0.21 & \textbf{0.69} & \textbf{26.48} & 0.39 & \underline{0.02}\\
VALL-E $[\spadesuit]$ \cite{valle} & LT (500) & \underline{3.61} & 0.21 & 0.24 & \underline{0.28} & 0.55 & 37.87 & 0.40 & \underline{0.02}\\
% StyleTTS 2 $[\bigstar]$ \cite{styletts2} & LT & 585 & - & - & - & - & - & - \\
% \rowcolor{verylightred}
% \rowcolor{verylightred}
% XTTS v2 & - & - & 3.54 & 0.07 & 0.38 & - & 1.27 & 1.44 \\
% E2-TTS $[\spadesuit]$ \cite{e2tts} & LT & 500 & ? & ? & ? & ? & ? & ? \\
\rowcolor{verylightgray}
\midrule
OZSpeech & LT (500) & 3.17 & \textbf{0.05} & 0.30 & \textbf{0.33} & \underline{0.62} & \underline{27.7} & 0.49 & \underline{0.02}\\
\midrule
% \rowcolor{verylightgray}
\multicolumn{10}{c}{\textit{3s Prompt}} \\
\midrule
% YourTTS $[\bigstar]$ \cite{yourtts} & LT + VCTK + T-P + M-A F & 640 & \underline{3.66} & \underline{0.08} & 0.34 & - & \textbf{1.09} & \underline{1.14} \\
F5-TTS $[\bigstar]$ \cite{f5tts} & E (95,000) & \textbf{3.76} & 0.24 & \textbf{0.53} & - & \underline{0.80} & 13.78 & \textbf{0.67} & \textbf{0.01} \\
VoiceCraft $[\bigstar]$ \cite{voicecraft} & GS (9,000) & 3.55 & 0.18 & \underline{0.51} & 0.45 & 0.78 & 17.22 & \underline{0.44} & \textbf{0.01}\\
NaturalSpeech 2 $[\clubsuit]$ \cite{ns2} & LT (585) & 2.38 & \underline{0.09} & 0.31 & 0.38 & \underline{0.80} & \underline{15.62} & 0.25 & \underline{0.02}\\
VALL-E $[\spadesuit]$ \cite{valle} & LT (500) & \underline{3.68} & 0.19 & 0.40 & \textbf{0.48} & 0.75 & 21.66 & 0.36 & \underline{0.02}\\
% \rowcolor{verylightred}
% \rowcolor{verylightred}
% XTTS v2 & - & - & 3.84 & 0.06 & 0.50 & - & 1.10 & 1.16 \\
% E2-TTS $[\spadesuit]$ \cite{e2tts} & LT & 500 & ? & ? & ? & ? & ? & ? \\
\rowcolor{verylightgray}
\midrule
OZSpeech & LT (500) & 3.15 & \textbf{0.05} & 0.40 & \underline{0.47} & \textbf{0.81} & \textbf{11.96} & \textbf{0.67} & \textbf{0.01}\\
\midrule
% \rowcolor{verylightgray}
\multicolumn{10}{c}{\textit{5s Prompt}} \\
\midrule
% YourTTS $[\bigstar]$ \cite{yourtts} & LT + VCTK + T-P + M-A F & 640 & \underline{3.64} & \underline{0.08} & \underline{0.48} & - & 1.17 & \underline{1.25} \\
F5-TTS $[\bigstar]$ \cite{f5tts} & E (95,000) & \textbf{3.72} & 0.32 & \textbf{0.58} & - & \underline{0.83} & \textbf{11.20} & \textbf{0.68} & \textbf{0.01} \\
VoiceCraft $[\bigstar]$ \cite{voicecraft} & GS (9,000) & \underline{3.58} & 0.19 & \underline{0.56} & \underline{0.51} & 0.81 & 14.48 & 0.46 & \textbf{0.01}\\
NaturalSpeech 2 $[\clubsuit]$ \cite{ns2} & LT (585) & 2.33 & \underline{0.09} & 0.35 & 0.44 & \textbf{0.84} & 13.13 & 0.28 & \underline{0.02}\\
VALL-E $[\spadesuit]$ \cite{valle} & LT (500) & \textbf{3.72} & 0.19 & 0.46 & \textbf{0.55} & 0.79 &  18.20 & 0.41 & \textbf{0.01}\\
% StyleTTS 2 $[\bigstar]$ \cite{styletts2} & LT & 585 & \textbf{4.20} & \textbf{0.03} & 0.42 & - & 1.11 & 1.17 \\
% \rowcolor{verylightred}
% \rowcolor{verylightred}
% XTTS v2 & - & - & 3.75 & 0.06 & 0.51 & - & 1.15 & 1.23 \\
% E2-TTS $[\spadesuit]$ \cite{e2tts} & LT & 500 & ? & ? & ? & ? & ? & ? \\
\rowcolor{verylightgray}
\midrule
OZSpeech & LT (500) & 3.15 & \textbf{0.05} & 0.39 & 0.48 & \underline{0.83} & \underline{12.05} & \underline{0.67} & \textbf{0.01} \\
\bottomrule
\end{tabular}
}
\end{center}
\vspace{-0.5cm}
\end{table*}

\subsection{Experiment Setup}
\textbf{Dataset}.
We employ the \textit{LibriTTS} dataset \cite{libritts}, which comprises multi-speaker English audio recordings of training data. 
For benchmarking purposes, we use the \textit{LibriSpeech test-clean} \cite{librispeech} dataset. 
More detailed information is provided in Appendix \ref{sec:data}.\\[8pt]
\textbf{Evaluation Metrics}.
To assess model performance, we employ the following objective evaluation metrics for each criterion: speech quality, quantified by UTMOS; speaker similarity, measured using SIM-O and SIM-R; robustness, indicated by WER; and prosody accuracy and error, analyzed through pitch and energy.
Additionally, we employ NFE and RTF metrics to measure the latency of the sampling process.
More details on the evaluation metrics can be found in Appendix \ref{sec:metrics}.\\[8pt]
\textbf{Baselines}. We compare our model with previous zero-shot TTS baselines. Further details regarding baselines are available in Appendix \ref{appendix:baselines}.

% \textbf{Baselines}. We compare our model with several zero-shot TTS baselines, including:
% \begin{itemize}
%     \item \textbf{YourTTS} \cite{yourtts}. We use the official code and pre-trained checkpoint \footnote{\url{https://github.com/Edresson/YourTTS}}, which is trained on LibriTTS, VCTK \cite{vctk}, TTS-Portuguese \cite{casanova2022tts}, and M-AILABS French datasets.
%     \item \textbf{VoiceCraft} \cite{voicecraft}. We use the official code and pre-trained checkpoint \footnote{\url{https://huggingface.co/pyp1/VoiceCraft/blob/main/830M_TTSEnhanced.pth}}, which is trained on the GigaSpeech dataset \cite{gigaspeech}.
%     \item \textbf{NaturalSpeech 2} \cite{ns2}. We use the Amphion toolkit \cite{amphion} and pre-trained checkpoint \footnote{\url{https://huggingface.co/amphion/naturalspeech2_libritts/tree/main/checkpoint}}, which is trained on the LibriTTS dataset.
%     \item \textbf{VALL-E} \cite{valle}. We reproduce VALL-E using the Amphion toolkit \cite{amphion} and train it under identical settings to our training dataset configuration.
% \end{itemize}

\subsection{Main Results}
\begin{table*}[!th]
\caption{Comparison of model size and latency for 3s audio prompt length. Column \textbf{\#Params} indicates the total number of parameters required for end-to-end synthesis, with the first value representing the parameters of the zero-shot model (trainable) and the second value corresponding to those of the neural codec or vocoder component (frozen).}
\label{tab:size-rtf-results}
\vspace{-0.5cm}
\begin{center}
\resizebox{\textwidth}{!}{%
\begin{tabular}{llcccc}
\toprule
\textbf{Model} & \textbf{\#Params} & \textbf{NFE} ($\downarrow$) & \textbf{RTF} ($\downarrow$) & \textbf{WER} ($\downarrow$) & \textbf{SIM-O} ($\uparrow$) \\
\midrule
% YourTTS \cite{yourtts} & 87M & - & \textbf{0.22} & \underline{0.09} & 0.34 \\

F5-TTS \cite{e2tts} & 336M + 13.5M Vovos \cite{vocos} & 32 & 0.70 & 0.24 & \textbf{0.53} \\
VoiceCraft \cite{voicecraft} & 830M + 14M EnCodec \cite{encodec} & - &  1.70 & 0.18 & \underline{0.51} \\
NaturalSpeech 2 \cite{ns2} & 378M + 14M EnCodec \cite{encodec} & 200 & 1.66 & 0.09 & 0.31 \\
VALL-E \cite{valle} & 594M + 104M SpeechTokenizer \cite{speechtokenizer} & - & 0.86 & 0.19 & 0.40 \\
\rowcolor{verylightgray}
\midrule
OZSpeech & 145M + 102M FACodec \cite{ns3} & \textbf{1} & \textbf{0.26} & \textbf{0.05} & 0.40 \\
\bottomrule
\end{tabular}
}
\end{center}
\vspace{-0.5cm}
\end{table*}

Table \ref{tab:overall-results} shows the performance of OZSpeech and representative baseline methods for 1s, 3s, and 5s audio prompt lengths.  OZSpeech establishes a new SOTA on WER across all audio prompt lengths, demonstrating superior content preserving capability through a \textit{multi-fold} reduction in WER. For example, OZSpeech reduces WER by a factor of $1.8 - 6.8$ over the other methods for 5s prompt length. Some of these models, such as F5-TTS, are trained on substantially more training data (F5-TTS is trained on 95,000 hours of speech whereas OZSpeech is trained on 500 hours). Compared to the next-best method, OZSpeech yields a relative reduction in WER by 58\%, 44\%, and 44\% for  1s, 3s, and 5s audio prompt lengths, respectively.  Additionally, while the WER scores of all baseline methods are sensitive to prompt length, OZSpeech maintains a consistent WER regardless of prompt length.

For pitch and energy accuracies and errors, which indicate the prosody reconstruction ability of TTS systems, OZSpeech consistently ranks as the best or second-best performer across different prompt lengths. For the remaining metrics (UTMOS, SIM-O, and SIM-R), our method in overall does not exhibit an obvious performance advantage over the baseline models. (We note that these models also experience trade-offs between different metrics.) However, our goal is to enhance the balance between intelligibility (i.e. content accuracy) and acoustical/perceptual quality while maintaining low latency and small model size.%However, our goal is to enhance the balance between intelligibility (i.e. content accuracy), model size, sampling speed, and speech synthesis capability without much compromise in speech perceptual qualities. % Pros: WER improvement is multifold. Sampling speed is fast. Model size is smallest

Our UTMOS scores show a rather small degradation compared to some of the baselines, particularly VALL-E and VoiceCraft. This is largely due to differences in the neural codecs' trade-offs between acoustic and semantic representations. EnCodec (VoiceCraft’s codec) primarily relies on acoustic codes, while SpeechTokenizer (VALL-E’s codec in this experiment) incorporates one semantic sequence alongside acoustic codes. In contrast, FaCodec (OZSpeech’s codec) strives to balance both representations. However, our focus is on optimizing the trade-off between sampling speed and speech synthesis quality.
We also retrain F5-TTS with 500~hours of the LibriTTS dataset using the official code for 1 million steps following its guideline, however the resulting WER exceeds 0.95 across all settings, so we exclude this retrained checkpoint from Table~\ref{tab:overall-results}  and instead use the officially released checkpoint, which is trained on 95,000 hours of data. The poor retraining results suggest that this method, which is based on traditional OT-CFM, requires a much larger, more diverse dataset for robustness. In contrast, neural codec-based models remain effective with limited data, likely due to extensive pre-training of the neural codec module on massive datasets. Thus, traditional OT-CFM methods like F5-TTS are unsuitable for low-resource languages.

Table \ref{tab:size-rtf-results} compares the model sizes and latency of OZSpeech and baseline models. OZSpeech is the smallest model of all, being only 29\%-71\% the size of the other models. When considering only the trainable part of the models, the number of trainable parameters of OZSpeech is only 17\%-43\% that of the other models. For the NFE metric, our model uses only a single sampling step, significantly reducing computation compared to NaturalSpeech 2 and F5-TTS, which require 200 and 32 steps, respectively, to achieve optimal performance. As a result, in terms of inference speed represented by the RTF metric, OZSpeech is almost 3 times faster than the next fastest model, F5-TTS.
%%%%%%%%%%%%%%%%%%%%%%%%%%%%%%%%%%%%%%%%%%%%%%%%%%%%%%%%%%%%%%%%%%%%%%%%%%%%%%%%%%%%%%%%%%%%%%%%%%%%%%%%%%%%%%%%%%%%%%%%%%%%%%%%%%%%
\subsection{Ablation Study}

\begin{table}
\caption{Comparison of two prompting strategies during training: \textit{First Segment} and \textit{Arbitrary Segment}.}
\label{tab:prompting-strategy}
\vspace{-0.4cm}
\begin{center}
\resizebox{\columnwidth}{!}{%
\begin{tabular}{lcccc}
\toprule
 & & & \multicolumn{2}{c}{\textbf{SPK-SIM}} 
 % & \multicolumn{2}{c}{\textbf{PESQ}} 
 \\
\cmidrule(lr){4-5} 
% \cmidrule(lr){6-7}
\textbf{Prompt Setting} & \textbf{UTMOS} ($\uparrow$) & \textbf{WER} ($\downarrow$) &\textbf{SIM-O} ($\uparrow$) & \textbf{SIM-R} ($\uparrow$)
% & \textbf{WB} ($\uparrow$)& \textbf{NB} ($\uparrow$)
\\
\midrule
\multicolumn{5}{c}{\textit{1s Prompt}} \\
\midrule
First segment & 3.01 & 0.08 & 0.25 & 0.29 
% & \textbf{1.14} & \textbf{1.28} 
\\
Arbitrary segment & \textbf{3.17} & \textbf{0.05} & \textbf{0.30} & \textbf{0.33} 
% & \textbf{1.14} & 1.27 
\\
\midrule
\multicolumn{5}{c}{\textit{3s Prompt}} \\
\midrule
First segment & 3.04 & 0.08 & 0.35 & 0.42 
% & 1.07 & 1.13 
\\
Arbitrary segment & \textbf{3.15} & \textbf{0.05} & \textbf{0.40} & \textbf{0.47} 
% & \textbf{1.08} & \textbf{1.15} 
\\
\midrule
\multicolumn{5}{c}{\textit{5s Prompt}} \\
\midrule
First segment & 3.02 & 0.06 & 0.37 & 0.45 
% & \textbf{1.20} & 1.26 
\\
Arbitrary segment & \textbf{3.15} & \textbf{0.05} & \textbf{0.39} & \textbf{0.48} 
% & \textbf{1.20} & \textbf{1.28} 
\\
\bottomrule
\end{tabular}
}
\end{center}
\vspace{-0.5cm}
\end{table}

Table \ref{tab:prompting-strategy} compares the performance of two prompting strategies: \textit{First Segment} and \textit{Arbitrary Segment}. The former generates prompts using the initial portion of the ground truth, whereas the latter selects random audio segments from the ground truth to form the prompts. The results clearly show that the \textit{Arbitrary Segment} strategy outperforms the \textit{First Segment} strategy across all metrics. In the \textit{First Segment} setting, the model seems to overfit in that it is forced to transfer the prompt to the beginning of the target. In contrast, the \textit{Arbitrary Segment} setting hides the position of the prompt, allowing it to smoothly transfer attributes from the prompt to the target. Consequently, we adopt the \textit{Arbitrary Segment} approach for our training.
This experiment also shows that the \textit{Arbitrary Segment} approach improves robustness by exposing the model to a more diverse range of speech contexts, leading to better generalization in zero-shot speech synthesis.

% Additional experimental results in the Appendix further validate the effectiveness of OZSpeech across different conditions. First, we conduct a computational efficiency analysis (Table \ref{tab:model-size-results}), demonstrating that OZSpeech requires significantly fewer parameters (only 145M for the trainable model) and achieves up to $6.5\times$ faster inference than competing methods, while maintaining comparable or superior content intelligibility. Second, we examine the impact of prompt length on synthesis quality (Fig.~\ref{fig:boxplot}), and find that OZSpeech maintains stable WER across varying prompt durations (1s, 3s, and 5s), unlike baseline models that show significant performance degradation with shorter prompts. This highlights the robustness of our model in low-resource scenarios. 

\subsection{Noise Tolerance Analysis}
\begin{table*}[!th]
\caption{Performance evaluation on noisy audio prompts. The noisy prompts are derived from the \textit{LibriSpeech test-clean} dataset with additive noise augmentation. The prompts are 3-second long. The checkpoints of each model trained on \textit{LibriTTS}—except for VoiceCraft, which was trained on \textit{GigaSpeech}—are used without re-training the models to include noisy samples. $[\blacklozenge]$ means fine-tuned results on noisy prompts. This table highlights a vulnerable use case where speech prompts contain noise, assessing the tolerance of these models. SNR$=\infty$ indicates the prompts are directly obtained from \textit{LibriSpeech test-clean} dataset and these results are simply copied from Table~\ref{tab:overall-results}.}
\label{tab:noisy-test}
\vspace{-0.5cm}
\begin{center}
\resizebox{\textwidth}{!}{%
\begin{tabular}{clcccccccc}
\toprule
 & & & & \multicolumn{2}{c}{\textbf{SPK-SIM}} & \multicolumn{2}{c}{\textbf{F0}} & \multicolumn{2}{c}{\textbf{Energy}} \\
\cmidrule(lr){5-6} \cmidrule(lr){7-8} \cmidrule(lr){9-10}
\textbf{SNR} (dB) & \textbf{Model} & \textbf{UTMOS} ($\uparrow$) & \textbf{WER} ($\downarrow$) &\textbf{SIM-O} ($\uparrow$) & \textbf{SIM-R} ($\uparrow$) & \textbf{Accuracy} ($\uparrow$) & \textbf{RMSE} ($\downarrow$)& \textbf{Accuracy} ($\uparrow$)& \textbf{RMSE} ($\downarrow$)\\
\midrule
\multirow{5}{*}{$\infty$} & F5-TTS \cite{f5tts} & \textbf{3.76} & 0.24 & \textbf{0.53} & - & \underline{0.80} & 13.78 & \textbf{0.67} & \textbf{0.01} \\
& VoiceCraft \cite{voicecraft} & 3.55 & 0.18 & \underline{0.51} & 0.45 & 0.78 & 17.22 & 0.44 & \textbf{0.01} \\
& NaturalSpeech 2 \cite{ns2} & 2.38 & 0.09 & 0.31 & 0.38 & 0.80 & 15.62 & 0.25 & \underline{0.02} \\
& VALL-E \cite{valle} & \underline{3.68} & 0.19 & 0.40 & \textbf{0.48} & 0.75 & 21.66 & 0.36 & \underline{0.02} \\
% \rowcolor{verylightred}
% \rowcolor{verylightred}
% & XTTS v2 & 3.30 & 0.07 & 0.46 & - & 1.10 & 1.19 \\
% & E2-TTS \cite{e2tts} & ? & ? & ? & ? & ? & ? \\
\cmidrule{2-10}
& \cellcolor{verylightgray} OZSpeech & \cellcolor{verylightgray} 3.15 & \cellcolor{verylightgray} \textbf{0.05} & \cellcolor{verylightgray} 0.39 & \cellcolor{verylightgray} \underline{0.47} & \cellcolor{verylightgray} \textbf{0.81} & \cellcolor{verylightgray} \textbf{11.96} & \cellcolor{verylightgray} \textbf{0.67} & \cellcolor{verylightgray} \textbf{0.01}\\

& \cellcolor{verylightgray} OZSpeech $[\blacklozenge]$ & \cellcolor{verylightgray} 3.19 & \cellcolor{verylightgray} \underline{0.06} & \cellcolor{verylightgray} 0.39 & \cellcolor{verylightgray} 0.46 & \cellcolor{verylightgray} 0.78 & \cellcolor{verylightgray} \underline{13.67} & \cellcolor{verylightgray} \underline{0.65} & \cellcolor{verylightgray} \textbf{0.01} \\

\midrule
\midrule
% \multirow{4}{*}{12} & YourTTS \cite{yourtts} & \textbf{3.60} & \underline{0.09} & \textbf{0.45} & - & 1.07 & 1.13 \\
% snr=12: {'utmos': 3.09165436721199, 'sim': {'SIM-O': 0.28796094699356956, 'SIM-R': 0.2778924810462959}, 'wer': 0.2504000282359273, 'prosody': {'F0 ACC': 0.4963621665319321, 'F0 RMSE': 42.20187310480414, 'ENERGY ACC': 0.4624090541632983, 'ENERGY RMSE': 0.028335905}}
\multirow{6}{*}{12} & F5-TTS \cite{f5tts} & \textbf{3.09} & 0.25 & 0.29 & - & 0.50 & 42.20 & 0.46 & 0.03 \\
& VoiceCraft \cite{voicecraft} & 2.42 & 0.20 & \textbf{0.40} & \textbf{0.40} & 0.59 & 32.48 & \underline{0.60} & \textbf{0.01} \\
& NaturalSpeech 2 \cite{ns2} & 1.66 & 0.12 & 0.22 & 0.34 & \underline{0.71} & \underline{20.0} & 0.45 & \textbf{0.01} \\
& VALL-E \cite{valle} & 2.43 & 0.51 & 0.25 & 0.31 & 0.54 & 59.25 & 0.40 & \underline{0.02}\\
% \rowcolor{verylightred}
% \rowcolor{verylightred}
% & XTTS v2 & 3.30 & 0.07 & 0.46 & - & 1.10 & 1.19 \\
% & E2-TTS \cite{e2tts} & ? & ? & ? & ? & ? & ? \\
\cmidrule{2-10}
& \cellcolor{verylightgray} OZSpeech & \cellcolor{verylightgray} 2.65 & \cellcolor{verylightgray} \textbf{0.05} & \cellcolor{verylightgray} 0.28 & \cellcolor{verylightgray} 0.35 & \cellcolor{verylightgray} 0.70 & \cellcolor{verylightgray} 21.70 & \cellcolor{verylightgray} 0.53 & \cellcolor{verylightgray} 0.03\\

& \cellcolor{verylightgray} OZSpeech $[\blacklozenge]$ & \cellcolor{verylightgray} \underline{3.04} & \cellcolor{verylightgray} \textbf{0.05} & \cellcolor{verylightgray} \underline{0.33} & \cellcolor{verylightgray} \underline{0.39} & \cellcolor{verylightgray} \textbf{0.76} & \cellcolor{verylightgray} \textbf{15.0} & \cellcolor{verylightgray} \textbf{0.73} & \cellcolor{verylightgray} \textbf{0.01} \\

\midrule
\midrule
% \multirow{4}{*}{6} & YourTTS \cite{yourtts} & \textbf{3.60} & \underline{0.08} & \textbf{0.44} & - & 1.07 & 1.12 \\
% snr=6: {'utmos': 2.41089639800437, 'sim': {'SIM-O': 0.29474728591039945, 'SIM-R': 0.30800200225303526}, 'wer': 0.2715409009974714, 'prosody': {'F0 ACC': 0.549717057396928, 'F0 RMSE': 40.94915536039594, 'ENERGY ACC': 0.5012126111560227, 'ENERGY RMSE': 0.11138506}}
\multirow{4}{*}{6} & F5-TTS \cite{f5tts} & \underline{2.41} & 0.27 & \underline{0.29} & - & 0.55 & 40.95 & 0.50 & 0.11\\
& VoiceCraft \cite{voicecraft} & 1.80 & 0.27 & \textbf{0.33} & \textbf{0.36} & 0.50 & 45.31 & \underline{0.68} & \textbf{0.01} \\
& NaturalSpeech 2 \cite{ns2} & 1.42 & 0.16 & 0.17 & 0.30 & \underline{0.61} & \underline{27.41} & 0.58 & \textbf{0.01} \\
& VALL-E \cite{valle} & 1.66 & 0.77 & 0.14 & 0.18 & 0.40 & 96.93 & 0.44 & \underline{0.02} \\
% \rowcolor{verylightred}
% \rowcolor{verylightred}
% & XTTS v2 & 2.98 & 0.09 & 0.41 & - & 1.10 & 1.19 \\
% & E2-TTS \cite{e2tts} & ? & ? & ? & ? & ? & ? \\
\cmidrule{2-10}
& \cellcolor{verylightgray} OZSpeech & \cellcolor{verylightgray} 2.21 & \cellcolor{verylightgray} \textbf{0.06} & \cellcolor{verylightgray} 0.23 & \cellcolor{verylightgray} 0.29 & \cellcolor{verylightgray} \underline{0.61} & \cellcolor{verylightgray} 32.80 & \cellcolor{verylightgray} 0.46 & \cellcolor{verylightgray} 0.05\\

& \cellcolor{verylightgray} OZSpeech $[\blacklozenge]$ & \cellcolor{verylightgray} \textbf{2.90} & \cellcolor{verylightgray} \textbf{0.06} & \cellcolor{verylightgray} \underline{0.29} & \cellcolor{verylightgray} \underline{0.34} & \cellcolor{verylightgray} \textbf{0.72} & \cellcolor{verylightgray} \textbf{17.41} & \cellcolor{verylightgray} \textbf{0.74} & \cellcolor{verylightgray} \textbf{0.01} \\
\midrule
\midrule
% \multirow{4}{*}{0} & YourTTS \cite{yourtts} & \textbf{3.57} & \underline{0.08} & \textbf{0.42} & - & 1.06 & 1.11 \\
\multirow{6}{*}{0} & F5-TTS \cite{f5tts} & \underline{1.88} & 0.32 & 0.17 & - & 0.41 & 51.80 & 0.44 & 0.29\\
& VoiceCraft \cite{voicecraft} & 1.58 & 0.44 & \underline{0.22} & \textbf{0.29} & 0.40 & 57.40 & \underline{0.55} & \underline{0.02} \\
& NaturalSpeech 2 \cite{ns2} & 1.33 & 0.23 & 0.12 & 0.26 & \underline{0.48} & \underline{38.27} & \textbf{0.56} & \textbf{0.01} \\
& VALL-E \cite{valle} & 1.44 & 0.93 & 0.07 & 0.11 & 0.36 & 102,68 & 0.52 & 0.07\\
% \rowcolor{verylightred}
% \rowcolor{verylightred}
% & XTTS v2 & 2.66 & 0.19 & 0.33 & - & 1.12 & 1.22 \\
% & E2-TTS \cite{e2tts} & ? & ? & ? & ? & ? & ? \\
\cmidrule{2-10}
& \cellcolor{verylightgray} OZSpeech & \cellcolor{verylightgray} 1.72 & \cellcolor{verylightgray} \textbf{0.06} & \cellcolor{verylightgray} 0.17 & \cellcolor{verylightgray} 0.22 & \cellcolor{verylightgray} 0.45 & \cellcolor{verylightgray} 46.60 & \cellcolor{verylightgray} 0.44 & \cellcolor{verylightgray} 0.08\\

& \cellcolor{verylightgray} OZSpeech $[\blacklozenge]$ & \cellcolor{verylightgray} \textbf{2.58} & \cellcolor{verylightgray} \textbf{0.06} & \cellcolor{verylightgray} \textbf{0.23} & \cellcolor{verylightgray} \underline{0.28} & \cellcolor{verylightgray} \textbf{0.67} & \cellcolor{verylightgray} \textbf{21.37} & \cellcolor{verylightgray} 0.54 & \cellcolor{verylightgray} \underline{0.02} \\
\bottomrule
\end{tabular}
}
\end{center}
\vspace{-0.5cm}
\end{table*}

 Unlike previous studies, we propose to investigate the effects of noisy prompts in TTS. Table \ref{tab:noisy-test} evaluates the tolerance of each model on noisy prompts. We conduct \textit{zero-shot} testing of all models in this scenario, where zero-shot in this context means that the models trained on their original training dataset, typically with clean prompts, are directly tested on noisy prompts. For this, we generate three sets of noisy prompts at SNRs of 0dB, 6dB, and 12dB, respectively. $\text{SNR} =\infty$ refers to prompts directly sourced from \textit{LibriSpeech test-clean} dataset, with metric values directly replicated from Table~\ref{tab:overall-results}.

Overall, all baseline methods are highly sensitive to noise in the audio prompts, experiencing significant degradation in all metrics as prompt SNR decreases. OZSpeech also shows similar sensitivity except for WER, which experiences either no or negligible degradation across all prompt SNR levels. VALL-E seems to be the most vulnerable to noise, where the WER increases by almost 2.7 times at the least noisy setting, SNR = 12dB. At SNR = 0dB, VALL-E  becomes almost unintelligible with a 93\% WER. The WER results highlight the robust intelligibility of OZSpeech, even in noisy prompt conditions. 

Although OZSpeech performs sub-optimally in non-WER metrics with the original clean prompts, it surpasses all baseline models in UTMOS. This improvement is largely attributed to the significant performance drop observed in the baseline models. Mixed results are observed when comparing all models on the remaining metrics.% (speaker similarity, pitch, and energy)
% The performance drop in OZSpeech when zero-shot on noisy prompts is much less than that of the baselines.

Next, we further fine-tune OZSpeech with both original and noisy prompts, where noisy prompts occur with a probability of 0.8. The noisy prompts are constructed by mixing the original prompts with random noise at different SNRs, drawn from a uniform distribution over the [0dB, 15dB] range. We leveraged the QUT-NOISE database \cite{noise} as our noise dataset. When tested with clean prompts (SNR = $\infty$), there is either no change or minimal changes in OZSpeech's performance across all metrics before and after fine-tuning. In noisy prompt conditions, WER remains unaffected by fine-tuning while the other metrics are significantly improved across all SNR levels. With decreasing SNR, fine-tuning generally yields increasingly larger improvements in all non-WER metrics.

We have empirically demonstrated the feasibility of the noise-aware training approach, which aims to synthesize noise-free speech conditioned by noisy prompts. This approach enables Zero-Shot TTS models to implicitly remove noise from given codec codes while preserving key attributes of speech. Consequently, neural codec-based Zero-shot TTS systems, which have traditionally been vulnerable and sensitive to noisy prompts, exhibit enhanced robustness, particularly against adversarial attacks.
\section{Conclusion}
We propose OZSpeech, an effective and efficient zero-shot TTS model that employs flow matching with a single sampling step from a learned prior instead of random noise. The model strikes a balance between synthesized speech intelligibility and acoustical quality. In particular, OZSpeech yields a multi-fold improvement in WER compared to existing baseline methods with some trade-off in the acoustical quality. Furthermore, unlike other methods, OZSpeech achieves a consistent WER across different audio prompt's lengths and noise levels. With a single-step sampling approach and a novel prior learning module that learns an effective starting point for the sampling process, our model requires significantly less computation, with inference speed being $2.7 - 6.5$ times faster than the other methods. In addition, our model size is only 29\%-71\% that of the other models. OZSpeech achieves competitive results even over models that are trained on much larger training sets.% These advantages form a a low-latency and efficient Zero=shot TTS system that 
% It outperforms existing SOTA methods with a significantly smaller model size and substantially faster inference times. 
 
In future work, we plan to enhance OZSpeech by integrating adaptive noise filtering techniques and expanding its capability to support multilingual and multimodal zero-shot speech synthesis, enabling more versatile applications in real-world scenarios.

\section*{Limitations} 
Despite achieving remarkable results, our Zero-shot TTS model still encounters challenges in naturalness. 
We have observed that the synthesized speech often exhibits slight distortions, which appear to contribute to a degradation in overall quality.
In this study, we employed the Duration Predictor, originally proposed in FastSpeech \cite{fastspeech}, to align input phonemes with codec codes. 
This module requires ground-truth phoneme durations for training; however, since phoneme durations are inherently real numbers rather than integers, inaccuracies arise in the ground-truth data. 
To align with codec codes, these durations must be rounded to integer values, which subsequently degrades the quality of the synthesized speech in the temporal domain.
To address this issue in future work, we plan to explore alternative alignment methods, such as Monotonic Alignment Search \cite{glowtts} or Encoder-Decoder architectures.
Nevertheless, the approach employed in this study remains the de facto approach in many real-world TTS systems, where latency is a critical factor.
Thus, this presents a trade-off between synthesis quality and computational efficiency.

\section*{Potential Risks}
Zero-shot Text-to-Speech (TTS) models offer several advantages, such as the ability to rapidly and effortlessly synthesize speech without requiring repeated recordings, making them particularly beneficial for content creators and for restoring damaged audio. However, despite these benefits, they also pose significant risks.
Zero-shot TTS models, which can generate speech in novel voices with little to no training data, present several potential threats, including:

\begin{compactitem}
    \item \textbf{Deepfake Fraud:} Malicious entities may exploit these models to impersonate individuals, facilitating scams, misinformation, or fraudulent activities.
    \item \textbf{Fabricated Media:} Synthesized audio can be used to create misleading or defamatory content, influencing public perception and spreading misinformation.
    \item \textbf{Privacy Violations:} The unauthorized replication of voices without explicit consent raises ethical and legal concerns regarding individual privacy.
    \item \textbf{Legal and Copyright Challenges:} Certain voices may be subject to copyright, trademark, or publicity rights protections, potentially leading to legal disputes over their unauthorized use.
\end{compactitem}


\bibliography{refs}

\appendix

% \section{Appendix}
\label{sec:appendix}
\section{Background}
\subsection{FACodec}
\label{appendix:FACodec}
Factorized neural speech codec, named FACodec, \cite{ns3} was proposed as a codec disentangler and timbre extractor. It separates the original speech waveform into distinct aspects: content, prosody, acoustic details, and timbre. Specifically, the speech input $x \in \mathbb{R}^C$ is processed through a speech encoder, $f_{enc}$, comprising several convolutional blocks to produce a pre-quantization latent representation:
\begin{equation}
    \label{eq:f_enc}
    h = f_{enc}(x) \in \mathbb{R}^{T \times D},
\end{equation}
where $T$ and $D$ denote the downsampled timeframes and the latent dimension, respectively. Subsequently, three factorized vector quantizers (FVQs) are employed to tokenize $h$ into distinct discrete sequences, capturing detailed representations of speech attributes such as content, prosody, and acoustic details. Let $\mathcal{Q}_p$, $\mathcal{Q}_c$, and $\mathcal{Q}_a$ denote the FVQs for prosody, content, and acoustic details, respectively. Each FVQ comprises a certain number of quantizers, defined $\mathcal{Q}_i = \{q_i^j\}^{N_i}_{j=1}$ where $i \in \{p,c, a\}$, $q_i^j \in \mathbb{R}^d$ represents the $j$-th quantizer corresponding to the $i$-th attribute, with a hidden dimension $d$, and its codebook size of 1024. The number of quantizers for each attribute is $N_p=1, N_c=2, N_a=3$. Thus, the output consists of a total of six sequences of discrete codes:
\begin{equation}
    \label{z}
    z = \textbf{Concat}(f_p(h), f_c(h), f_a(h)) \in \mathbb{R}^{T \times 6},
\end{equation}
where $f_p(h) \in \mathbb{R}^{T \times 1}$, $f_c(h) \in \mathbb{R}^{T \times 2}$, and $f_a(h) \in \mathbb{R}^{T \times 3}$ are functions that map the latent representation $h$ into discrete codes representing the speech attributes, which are then concatenated into a unified representation $z$.

The timbre attribute is extracted by passing $h$ through several Conformer blocks \cite{gulati20_interspeech} combined with a temporal pooling layer, which converts $h$ into a timbre-specific representation:
\begin{equation}
    \label{z_t}
    z_t = \textbf{TemporalPooling}(\textbf{Conformer}(h)) \in \mathbb{R}^D.
\end{equation}
After obtaining $z$ and $z_t$, the neural codec decoder $f_{dec}$ combines them to reconstruct the waveform:
\begin{equation}
    \label{eq:f_dec}
    y = f_{dec}(z, z_t).
\end{equation} 
Inspired by Eq.~\eqref{eq:f_dec}, which takes $z$ and $z_t$ as inputs and is pre-trained on a large-scale, multi-speaker dataset, ensuring robust zero-shot TTS capabilities, our approach aims to build a system that generates a six-sequence representation $\tilde{z} \in \mathbb{R}^{T \times 6}$, which is forced to lie within the subspaces of the pre-trained FACodec. This representation captures prosody, content, and acoustic details in a manner consistent with $z$. Subsequently, $\tilde{z}$ is fed into $f_{dec}$, alongside $z_t$, obtained using Eq.~\eqref{z_t}, to synthesize the speech output $\tilde{y}$.

\subsection{Flow Matching}
\label{appendix:fm}
We present the fundamental principles of Flow Matching (FM) upon which our model is built. FM aims to construct a probability path $x_t \sim p_t(x)$, from a known source distribution $x_0 \sim p_0(x)$ (typically a Gaussian distribution) to a target distribution $x_1 \sim p_1(x)$. Specifically, FM is formulated as a regression objective for training a velocity field (also called a vector field), which models the instantaneous velocities of samples at time $t$ (also known as the flow). This velocity field is then used to transform the source distribution $p_0$ into the target distribution $p_1$ along the probability path $p_t$. Formally, the flow of $x$ along the trajectory is defined by an ordinary differential equation (ODE):
\begin{equation}
    \label{eq:ode_flow}
    \frac{d}{dt}d\psi_t(x) = \mathbf{v}_{t}(\psi_t(x);\theta),\quad \psi_0(x) = x,
\end{equation}
where $t \sim \mathcal{U}[0,1]$, $\psi_t: [0,1] \times \mathbb{R}^d \to \mathbb{R}^d$ represents a time-dependent flow describing the position of the point $x$ at time $t$, and $\mathbf{v}_{t}: [0,1] \times \mathbb{R}^d \to \mathbb{R}^d$ is the time-dependent velocity field modeled by a neural network with parameters $\theta$. Given $x_t := \psi_t(x_0)$, the velocity field $\mathbf{v}_t$ creates a probability path $p_t$ such that $x_t \sim p_t$ for $x_0 \sim p_0$. Under this formulation, the objective is to regress velocity field $\mathbf{v}_t$ predicted by the neural network parameterized by $\theta$ to a target velocity field $\mathbf{u}_t$ in order to generate the desired probability path $p_t$. This is achieved by minimizing the Flow Matching (FM) loss:
\begin{equation}
\label{eq:fm}
\begin{split}
    \mathcal{L}_{FM}(\theta) = \mathbb{E}_{t,x_t}
    \Big\|
        \mathbf{v}_t(x_t;\theta) - \mathbf{u}_t(x_t)
    \Big\|^2,
\end{split}
\end{equation}
where $t \sim \mathcal{U}[0,1],~x_t \sim p_t$.

In practice, $\mathcal{L}_{FM}(\theta)$ is rarely implemented due to the complexity of $\mathbf{u}_t$ and the lack of prior knowledge of $p_t$, $\mathbf{u}_t$, and the target distribution $p_1$, which makes it an obstacle to directly calculate $\mathbf{u}_t(x_t)$. A feasible approach to address this issue is to simplify the loss by constructing the probability path $p_t$ conditioned on real data $x_1$ from the training dataset. This path is also known as \textit{conditional optimal transport} path. Following \cite{ot-cfm}, a random variable $x_t \sim p_t$ can be expressed as a linear combination of $x_0 \sim \mathcal{N}(x |0,I)$ and $x_1 \sim p_1$:
\begin{equation}
    x_t = tx_1 + (1-t)x_0 \sim p_t,
\end{equation}
Thus, the probability path $p_t(x|x_1)=\mathcal{N}(x|tx_1,(1-t)^2I)$. Given $x_t$ represents conditional random variables, the conditional velocity field can be derived from $\frac{d}{dt}x_t = \mathbf{u}_t(x_t|x_1)$ as $\mathbf{u}_t(x_t|x_1) = x_1 - x_0$. Using this, we can formulate a tractable and simplified version of the Flow Matching loss \eqref{eq:fm}, referred to as the Conditional Flow Matching (CFM) loss. This formulation encourages straighter trajectories between the source and target distributions and is expressed as follows:
\begin{equation}
\label{eq:orgin-cfm}
\begin{split}
    \mathcal{L}_{CFM}(\theta)   & = \mathbb{E}_{t,x_0,x_1}
    \Big\|
        \mathbf{v}_t(x_t;\theta) - \mathbf{u}_t(x_t|x_1)
    \Big\|^2 \\
        & = \mathbb{E}_{t,x_0,x_1}
    \Big\|
        \mathbf{v}_t(x_t;\theta) - (x_1 - x_0)
    \Big\|^2,
\end{split}
\end{equation}
where $t \sim \mathcal{U}[0,1],~x_0 \sim \mathcal{N}(x |0,I),~x_1 \sim p_1$. Once the training of the vector field $\mathbf{v}_t$ is complete, solving the ODE \eqref{eq:ode_flow} at discretized time steps until $t=1$ allows us to generate novel samples $x_1$ that approximate the target distribution $p_1$.

\section{Method Details}
\label{sec:training}

\textbf{Prompting Trick during Training}.
As outlined in Section \ref{sec:osfm}, incorporating an acoustic prompt is essential for generating $\textbf{x}_1$. This process involves transferring prosody and acoustic detail attributes from the prompt to the output quantizers. 
A significant challenge arises in preparing prompt-target pairs that exhibit similar attributes, as mismatches can lead to degraded performance.
To address this issue, we leverage ground truth quantizers, utilizing them as both the prompt and the target during training. 
Specifically, we randomly select and clone a segment of 1 $\sim$ 3 seconds from the ground truth data to serve as the prompt at each training step. 
This approach ensures a high degree of similarity between the prompt and target, facilitating more effective attribute transfer and enhancing the quality of the generated output.\\[8pt]
\textbf{Losses Computing Strategy}.
Let the velocity of $\breve{\mathbf{z}}_{\text{pr}}$ along the path progressing toward the corresponding $\mathbf{z}_1$ be represented as $\mathbf{v}_{\theta}^{1:L}(\breve{\mathbf{z}}_{\text{pr}}, \tau)$, where $L$ denotes the length of the entire output sequence of $\mathbf{v}_{\theta}(\cdot,\cdot)$.
Our goal is to compute the drift of $\mathbf{x}_{\text{pr}}$ for generating $\mathbf{x}_1$ only; it is unnecessary to backpropagate gradients over the entire output sequence, which includes the concatenation of acoustic prompt $\textbf{y}$ and $\textbf{x}_{\text{prior}}$ velocities.
To address this, $\mathbf{v}_{\theta}^{1:L}(\breve{\mathbf{z}}_{\text{pr}}, \tau)$ is truncated by excluding the velocity components associated with the acoustic prompt $\mathbf{y}$, where $M$ denotes its length.
The resulting truncated velocity, $\mathbf{v}_{\theta}^{M:L}(\breve{\mathbf{z}}_{\text{pr}}, \tau)$, is then used in subsequent operations, including loss computation, to ensure computational efficiency while maintaining the focus on the target velocity for $\mathbf{x}_{\text{pr}}$.
As a result, Eq. (\eqref{eq:ot-cfm-adjusted-1}) is rewritten as:
\begin{equation}
    \label{eq:ot-cfm-adjusted-2}
    \mathcal{L}_{CFM}(\theta) = \mathbb{E}_{\mathbf{x}_1, \mathbf{x}_{\text{pr}}}
    \left\lVert
        \mathbf{v}_{\theta}^{M:L}(\breve{\mathbf{z}}_{\text{pr}}, \tau) - \frac{\mathbf{x}_1 - \mathbf{x}_{\text{pr}}}{1 - \tau}
    \right\rVert^2.
\end{equation}
Consequently, the Anchor Loss $\mathcal{L}_{anchor}$ approximating the target distribution $\mathbf{x}_1$, conditioned on the $\tilde{\mathbf{x}}_1$ is formulated:
\begin{equation}
    \label{eq:anchor-loss-2}
    \mathcal{L}_{anchor}(\phi) = \mathbb{E}_{\mathbf{x}_1, \tilde{\mathbf{x}}_{1}} \left[ - \mathbf{log} p(\mathbf{x}_1 \mid \tilde{\mathbf{x}}_1;\phi) \right],
\end{equation}
where, $\tilde{\mathbf{x}}_1$ is computed as follows:
\begin{equation*}
    \label{eq:x1-est-2}
    \tilde{\mathbf{x}}_1 = \mathbf{x}_{\text{pr}} + (1 - \tau)\mathcal{F}^{-1}(\mathbf{v}_{\theta,M:L}(\breve{\mathbf{z}}_{\text{pr}}, \tau)).
\end{equation*}

\section{Training Details}

We integrate the Prior Codes Generator and the Vector Field Estimator, exploring various configurations to optimize overall system performance. 
For the Prior Codes Generator, we employ a compact neural network architecture with the following specifications: a hidden dimension of $d_{\text{model}} = 256$, a multi-head attention mechanism with $n_{\text{heads}} = 4$, and a feed forward network filter size of $d_{\text{ffn}} = 1024$.
These parameters are consistently applied across both the encoder and decoder layers.
The architecture includes 2 FFT blocks in both the encoder and the shared decoder, while an additional 6 blocks are utilized as specific layers to estimate the corresponding quantizers. 
The output dimensionality of the Prior Codes Generator is set to $\mathbf{x}_{\text{pr}} = 1024$, ensuring alignment with subsequent processing stages.
For the Vector Field Estimator, we adopt a Transformer architecture comprising four layers, each characterized by a hidden dimension of $d_\text{modlel} = 1024$, a number of attention heads $n_\text{heads} = 32$, and a feedforward network inner dimension of $d_\text{ffn} = 4096$ for the base-size model. 

The Prior Codes Generator and the Vector Field Estimator are jointly trained on a cluster of four 80GB A100 GPUs, using a batch size of $16$.
The training process employs the \textit{AdamW} optimizer with a learning rate of $10^{-4}$, $\beta_1 = 0.9$, $\beta_2 = 0.98$, and a weight decay parameter of $10^{-4}$. 
% Additionally, a \textit{Cosine Annealing Scheduler} is utilized, incorporating $5000$ warm-up steps and $50.000$ decay steps.

% \textbf{\emph{Do not put content after the references.}}

    % \begin{figure*}[htbp]
    %     \centering
    %     \includegraphics[width=1\textwidth]{figs/x0-virtual.pdf}
    %     \caption{$x_0$ virtual}
    %     \label{fig:x0-vitual}
    % \end{figure*}

\section{Evaluation Details}
\subsection{Dataset Details}
\label{sec:data}
\textbf{Training dataset}. We use a subset of 500 hours from the \textit{LibriTTS} dataset, where the duration of individual audio ranges from 1.0 to 16.6 seconds.
From this dataset, we construct metadata for each training sample, which includes the following elements: input phonemes, target durations, and target code sequences. 
To derive the input phonemes and their corresponding target durations, we use the Montreal Forced Alignment (MFA) \cite{mfa} tool. 
This tool aligns each audio sample with its transcription and extracts the duration of each phoneme.
Furthermore, we produce target codes using FACodec, which processes input waveforms sampled at 16 kHz. 
The FACodec applies a folding operation at a compression factor of 200. 
As a result, each second of audio is decomposed into a set of six quantizers, with each quantizer comprising 80 discrete speech units.
These units have a value range spanning from 0 to 1023.\\[8pt]
\textbf{Evaluation dataset}. We follow the VALL-E evaluation protocol \cite{valle}. Particularly, the LibriSpeech test-clean dataset is filtered to include samples between 4 and 10 seconds in length, totaling 2.2 hours of audio. For each sample, the prompt speech is randomly selected from another sample by extracting a 1-second, 3-second, or 5-second clip, depending on the prompt setting used in our experiment, from the same speaker.

\begin{table*}
\caption{Comparison of two OZSpeech model sizes: Base (145M parameters) and Small (100M parameters), evaluated on the \textit{LibriSpeech test-clean} dataset. Both models were trained on the 500-hour \textit{LibriTTS} training dataset.}
\label{tab:model-size-results}
\begin{center}
\resizebox{1.0\textwidth}{!}{%
\begin{tabular}{lcccccccc}
\toprule
 & & & \multicolumn{2}{c}{\textbf{SPK-SIM}} 
 % & \multicolumn{2}{c}{\textbf{PESQ}} 
 & \multicolumn{2}{c}{\textbf{F0}} & \multicolumn{2}{c}{\textbf{Energy}} \\
\cmidrule(lr){4-5} \cmidrule(lr){6-7} \cmidrule(lr){8-9} 
% \cmidrule(lr){10-11}
\textbf{Model Size} & \textbf{UTMOS} ($\uparrow$) & \textbf{WER} ($\downarrow$) &\textbf{SIM-O} ($\uparrow$) & \textbf{SIM-R} ($\uparrow$)
% & \textbf{WB} ($\uparrow$)& \textbf{NB} ($\uparrow$) 
& \textbf{Accuracy} ($\uparrow$) & \textbf{RMSE} ($\downarrow$) & \textbf{Accuracy} ($\uparrow$) & \textbf{RMSE} ($\downarrow$) \\
\midrule
% \rowcolor{verylightgray}
\multicolumn{9}{c}{\textit{1s Prompt}} \\
\midrule
Base & \textbf{3.17} & 0.05 & \textbf{0.30} & 0.33 
% & 1.14 & 1.27 
& 0.62 & 27.70 & 0.49 & 0.02 \\
Small & 3.15 & 0.05 & 0.29 & 0.33 
% & 1.14 & 1.27 
& \textbf{0.69} & \textbf{23.94} & \textbf{0.51} & 0.02 \\
\midrule
% \rowcolor{verylightgray}
\multicolumn{9}{c}{\textit{3s Prompt}} \\
\midrule
Base & \textbf{3.15} & \textbf{0.05} & \textbf{0.40} & \textbf{0.47} 
% & 1.08 & 1.15 
& \textbf{0.81} & \textbf{11.96} & \textbf{0.67} & 0.01 \\
Small & 3.14 & 0.06 & 0.37 & 0.44 
% & 1.08 & 1.15 
& 0.78 & 13.54 & 0.65 & 0.01 \\
\midrule
% \rowcolor{verylightgray}
\multicolumn{9}{c}{\textit{5s Prompt}} \\
\midrule
Base & 3.15 & 0.05 & \textbf{0.39} & \textbf{0.48} 
% & \textbf{1.20} & \textbf{1.28} 
& \textbf{0.83} & \textbf{12.05} & \textbf{0.67} & 0.01 \\
Small & \textbf{3.17} & 0.05 & 0.38 & 0.46 
% & 1.19 & 1.26 
& 0.79 & 12.58 & 0.66 & 0.01 \\
\bottomrule
\end{tabular}
}
\end{center}
\end{table*}
\subsection{Metrics Details}
\label{sec:metrics}
We evaluate each system using the following objective evaluation metrics:
\begin{itemize}
    \item \textbf{RTF} (Real-Time Factor) is an essential metric for assessing a system's efficiency, particularly in scenarios demanding real-time processing. It represents the time required to produce one second of speech. We assess the RTF of all models in a fully end-to-end setup using an NVIDIA 80GB A100 GPU.

    \item \textbf{NFE} (Number of Function Evaluations) denotes the total number of times the model's guiding function—often a score or drift function—is computed during the sampling process. This metric is especially important in settings where the generative process is formulated as solving an ordinary differential equation (ODE), such as in the probability flow ODE method used in score-based generative models.
    
    \item \textbf{UTMOS} \cite{utmos} is a deep learning-based system used to evaluate speech quality by predicting the mean opinion scores (MOS). It eliminates the need for costly, time-consuming subjective evaluations by using advanced deep learning techniques to provide predictions that closely align with human judgments.
    
    \item \textbf{SIM-O} and \textbf{SIM-R} are metrics used to evaluate speaker similarity. \textbf{SIM-O} measures the similarity between the synthesized speech and the original prompt, while \textbf{SIM-R} evaluates the similarity between the synthesized speech and the reconstructed prompt generated by FACodec \cite{ns3}. These metrics are computed by calculating the cosine similarity of speaker embeddings extracted by applying WavLM-TDCNN \footnote{\url{https://github.com/microsoft/UniSpeech/tree/main/downstreams/speaker_verification}} on the audio waveforms. Both SIM-O and SIM-R range from -1 to 1, with higher values indicating greater speaker similarity.
    
    % \item \textbf{PESQ} \cite{pesq} (Perceptual Evaluation of Speech Quality) is used to evaluate the perceptual quality. It compares the synthesized speech against the prompt speech signal by simulating human auditory perception. It combines time- and frequency-domain models to quantify distortions that affect perceived speech quality. PESQ generates a score on a scale of 1.0 (poor quality) to 4.5 (excellent quality), with higher values indicating better perceptual quality.
    
    \item \textbf{WER} (Word Error Rate) is used to evaluate the robustness of speech synthesis systems, specifically how accurately they pronounce each word. We employ an ASR model \footnote{\url{https://huggingface.co/facebook/hubert-large-ls960-ft}} to transcribe the generated speech and compare the transcription with the text prompt. The ASR model used is a CTC-based HuBERT, pre-trained on LibriLight and fine-tuned on the 960-hour training set of LibriSpeech.
    
    \item \textbf{Prosody Accuracy \& Error} are used to assess the alignment between the synthesized speech and audio prompt, with a specific focus on pitch (F0) and energy. For accuracy assessment, we adopt the methodology proposed in PromptTTS \cite{prompttts} and TextrolSpeech \cite{textrolspeech}, categorizing the F0 and energy levels of speech into three categories—high, normal, and low—based on their mean values \footnote{\url{https://github.com/jishengpeng/TextrolSpeech}}. Additionally, we employ the Root Mean Square Error (RMSE) to quantify the differences in F0 and energy between the synthesized speech and the corresponding prompts.
\end{itemize}
% \textbf{OZSpeech + HiFiGAN/BigVGAN}
\subsection{Baselines Details}
\label{appendix:baselines}
We compare our model with previous zero-shot TTS baselines, including:
\begin{itemize}
    \item \textbf{VoiceCraft} \cite{voicecraft}. We use the official code and pre-trained checkpoint \footnote{\url{https://huggingface.co/pyp1/VoiceCraft/blob/main/830M_TTSEnhanced.pth}}, which is trained on the GigaSpeech dataset \cite{gigaspeech}.
    \item \textbf{NaturalSpeech 2} \cite{ns2}. We use the Amphion toolkit \cite{amphion} and pre-trained checkpoint \footnote{\url{https://huggingface.co/amphion/naturalspeech2_libritts/tree/main/checkpoint}}, which is trained on the LibriTTS dataset \cite{libritts}.
    \item \textbf{F5-TTS} \cite{f5tts}. We use the official code and pre-trained checkpoint \footnote{\url{https://huggingface.co/SWivid/F5-TTS/blob/main/F5TTS_Base_bigvgan/model_1250000.pt}}, which is trained on the Emilia dataset \cite{emilia}.
    \item \textbf{VALL-E} \cite{valle}. We reproduce VALL-E using the Amphion toolkit \cite{amphion} and train it under identical settings to our training dataset configuration.
\end{itemize}

\section{Extra Experiments}
% \subsection{OZSpeech Variants}
Table~\ref{tab:model-size-results} shows the performance of OZSpeech-Base (145M parameters) and OZSpeech-Small (100M parameters). Although at over 31\% reduction in size, the Small model shows comparable performance with the Base model across all metrics, except for pitch (\textbf{F0}). Interestingly, the Small model outperforms the Base model by 13.6\% in F0 RMSE for 1s prompt length (23.94 for Small vs. 27.70 for Base). However, for 3s prompt length, it experiences a 13.2\% relative decline in the same metric (13.54 for Small vs. 11.96 for Base).

\section{Analysis}
\begin{figure*}[htbp]
    \centering
    \begin{subfigure}[b]{\textwidth}
        \includegraphics[width=\textwidth]{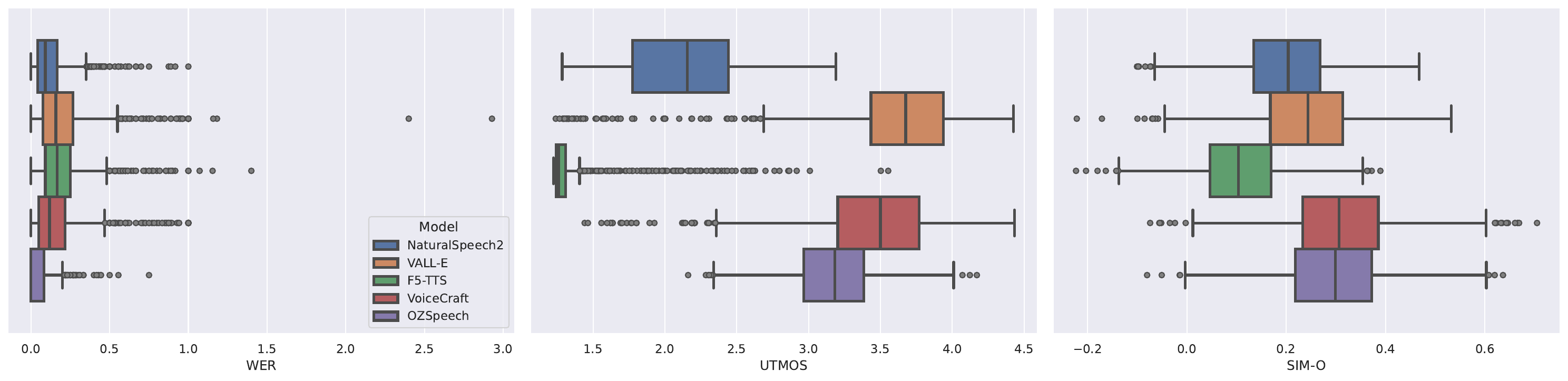}
        \caption{1s Prompt}
    \end{subfigure}
    \begin{subfigure}[b]{\textwidth}
        \includegraphics[width=\textwidth]{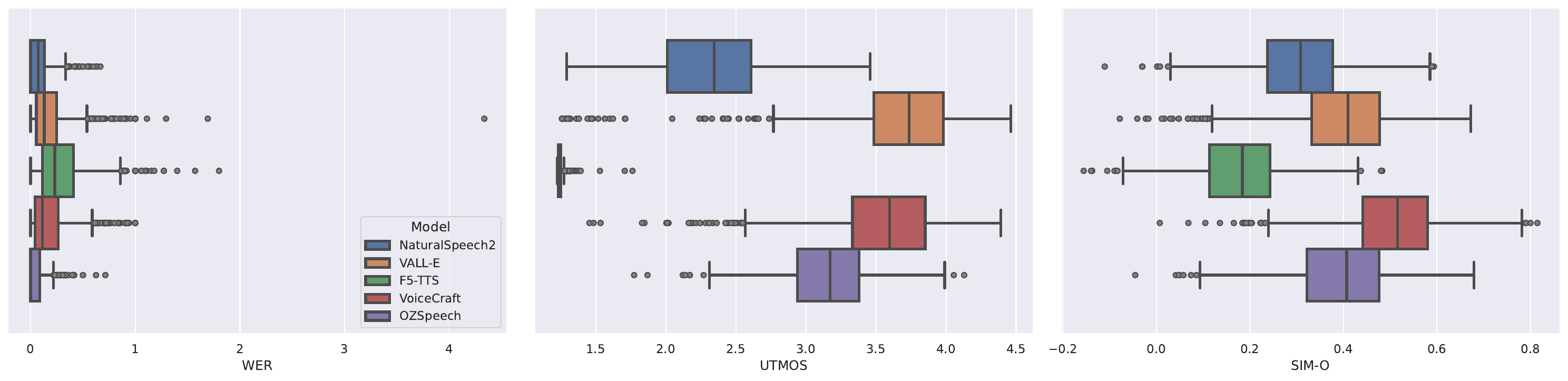}
        \caption{3s Prompt}
    \end{subfigure}
    \begin{subfigure}[b]{\textwidth}
        \includegraphics[width=\textwidth]{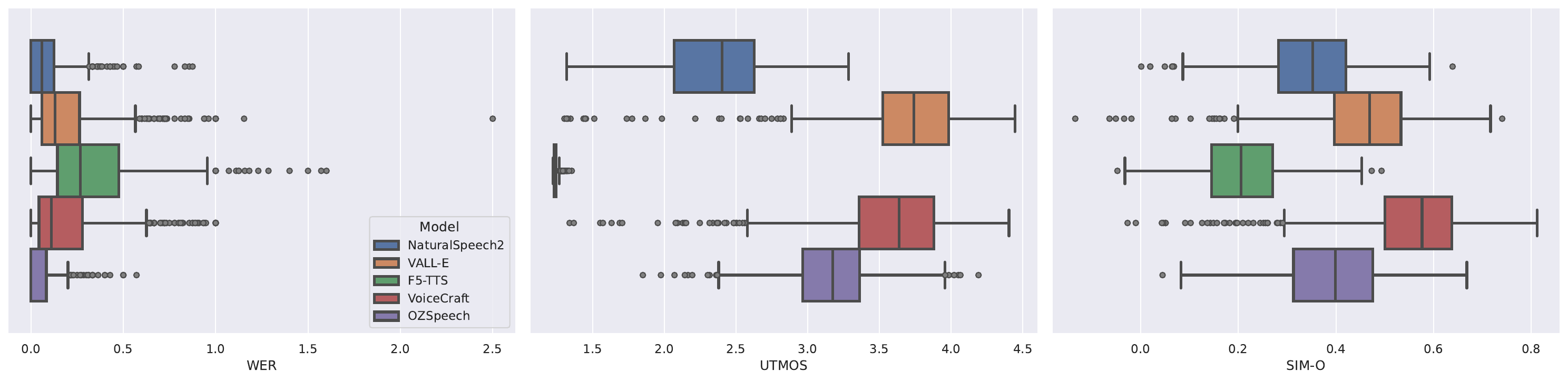}
        \caption{5s Prompt}
    \end{subfigure}
    \caption{Boxplots showing the distributions of performance metrics (WER, UTMOS, and SIM-O) on the LibriSpeech test-clean dataset for each model, evaluated across different audio prompt lengths.}
    \label{fig:boxplot}
\end{figure*}

As shown in Figure \ref{fig:boxplot}, the distributions of performance metrics across different prompt lengths are as follows:
\begin{itemize}
    \item \textbf{WER}: For the 1s prompt, OZSpeech exhibits a distribution that is very close to zero with a narrow box, indicating superior performance. The second best is NaturalSpeech 2, which shows a slightly right-shifted box compared to OZSpeech, followed by VoiceCraft. This pattern is consistent across the 3s and 5s prompts. In contrast, VALL-E and F5-TTS display higher distributions. Notably, F5-TTS shows slightly better performance than VALL-E for the 1s prompt. However, as the prompt length increases, F5-TTS significantly lags behind the other baselines, with its distribution approaching 0.5.
    \item \textbf{UTMOS}: The best performance for this metric is achieved by VALL-E for the 1s prompt. Its distribution shifts slightly to the right for the 3s prompt and stabilizes for the 5s prompt. VoiceCraft and OZSpeech show the next best performances, maintaining stable distributions across different prompt lengths, with VoiceCraft consistently outperforming OZSpeech. NaturalSpeech 2 scores mostly below 3.0 for the 1s prompt and shows improvement as the prompt length increases. Notably, F5-TTS consistently scores below 1.5, significantly lagging behind the other baselines.
    \item \textbf{SIM-O}: For the 1s prompt, the distributions of VoiceCraft and OZSpeech are almost equivalent, followed by VALL-E, F5-TTS, and NaturalSpeech 2, respectively. As the prompt length increases, the differences among the models become more noticeable. Specifically, VoiceCraft shows the best performance in retaining the speaker's identity in the synthesized output. VALL-E follows with the second-best performance, followed by OZSpeech and NaturalSpeech 2, respectively. In contrast, F5-TTS consistently demonstrates poor performance regardless of the prompt length.
\end{itemize}

All in all, OZSpeech achieved competitive performance compared to other baselines. Although our method does not show a clear performance advantage over baseline models, our primary objective is to balance sampling speed and speech synthesis capability. This trade-off enables our method to maintain a small model size with low WER while preserving the naturalness of the speech and the speaker's identity style from the prompt in the synthesized output. In contrast to the baselines, which have larger models and require longer inference times (see Table \ref{tab:size-rtf-results}), our method demonstrates a significant advantage. With just one sampling step, we can achieve promising performance.

\end{document}